\documentclass{aa}  
\usepackage{amsmath}
\usepackage[colorinlistoftodos]{todonotes}
\usepackage{natbib}
\bibpunct{(}{)}{;}{a}{}{,} % to follow the A&A style
\usepackage[T1]{fontenc}
\usepackage[english]{babel} 
\usepackage{amssymb}
\usepackage{mathrsfs}
\usepackage{soul} 
\usepackage{ulem} 
\usepackage{multicol,float}
\usepackage{caption}%
\usepackage{textcomp} 
\usepackage[pdftex,colorlinks=true,linkcolor=black,citecolor=blue,urlcolor=blue]{hyperref}
\everymath{\displaystyle}
\usepackage{tabularx}
\newcolumntype{C}{>{\centering}X}
\usepackage[version=4]{mhchem}

\usepackage{subcaption}
\usepackage{gensymb}
\usepackage{graphicx}
\usepackage{upgreek}

\usepackage{txfonts}

\usepackage{blindtext}

\makeatletter
\renewcommand{\paragraph}{\@startsection{paragraph}{4}{0ex}%
	{-3.25ex plus -1ex minus -0.2ex}%
	{1.5ex plus 0.2ex}%
	{\normalfont\normalsize}}
\makeatother

\stepcounter{secnumdepth}
\stepcounter{tocdepth}
\usepackage[switch, modulo]{lineno}
\modulolinenumbers[1]
\begin{document}

	\title{PDRs4All \\
    XI. Detection of infrared CH$^+$ and CH$_3^+$ rovibrational emission in the Orion Bar and disk d203-506: evidence of chemical pumping.}
 \titlerunning{Evidence of chemical pumping of CH$^+$ and CH$^+_3$ in the Orion Bar}

	\author{M. Zannese
		\inst{1}, B. Tabone\inst{1}, E. Habart\inst{1}, E. Dartois\inst{2}, J. R. Goicoechea\inst{3}, L. Coudert\inst{2}, B. Gans\inst{2}, M.-A. Martin-Drumel\inst{2}, U. Jacovella\inst{2}, A. Faure\inst{4}, B. Godard\inst{5}, A.G.G.M. Tielens\inst{6,7}, R. Le Gal\inst{4,8}, J. H. Black\inst{9}, S. Vicente\inst{10}, O. Berné\inst{11}, E. Peeters\inst{12,13,14}, D. Van De Putte\inst{12,13,15}, R. Chown\inst{12,13}, A. Sidhu\inst{12,13},  I. Schroetter\inst{11}, A. Canin\inst{11}, O. Kannavou\inst{1}}
	\authorrunning{M. Zannese
		\inst{1}}

	\institute{Université Paris-Saclay, CNRS, Institut d'Astrophysique Spatiale, 91405 Orsay, France\\
		\email{marion.zannese@universite-paris-saclay.fr} \and Universit\'e Paris-Saclay, CNRS, Institut des Sciences Mol\'eculaires d'Orsay, 91400 Orsay, France \and Instituto de F\'{\i}sica Fundamental  (CSIC),  Calle Serrano 121-123, 28006, Madrid, Spain \and Institut de Plan\'etologie et d'Astrophysique de Grenoble (IPAG), Universit\'e Grenoble Alpes, CNRS, F-38000 Grenoble, France \and Observatoire de Paris, Université PSL, Sorbonne Université, LERMA, 75014 Paris, France \and Leiden Observatory, Leiden University, 2300 RA Leiden, The Netherlands \and Astronomy Department, University of Maryland, College Park, MD 20742, USA \and Institut de Radioastronomie Millim\'etrique (IRAM), 300 Rue de la Piscine, F-38406 Saint-Martin d'H\`{e}res, France \and Department of Space, Earth and Environment, Chalmers University of Technology, Onsala Space Observatory, 43992, Onsala, Sweden \and Instituto de Astrof\'isica e Ci\^{e}ncias do Espa\c co, Tapada da Ajuda, Edif\'icio Leste, 2\,$^{\circ}$ Piso, P-1349-018 Lisboa, Portugal \and Institut de Recherche en Astrophysique et Plan\'etologie, Universit\'e Toulouse III - Paul Sabatier, CNRS, CNES, 9 Av. du colonel Roche, 31028 Toulouse Cedex 04, France \and Department of Physics \& Astronomy, The University of Western Ontario, London ON N6A 3K7, Canada \and Institute for Earth and Space Exploration, The University of Western Ontario, London ON N6A 3K7, Canada \and Carl Sagan Center, SETI Institute, 339 Bernardo Avenue, Suite 200, Mountain View, CA 94043, USA \and Space Telescope Science Institute, 3700 San Martin Drive, Baltimore, MD, 21218, USA}
	
	%\date{08/2022}
	
	% \abstract{}{}{}{}{} 
	% 5 {} token are mandatory
	
	\abstract
	{The methylidyne cation (CH$^+$) and the methyl cation (CH$_3^+$) are building blocks of organic molecules in the ultraviolet (UV)-irradiated gas, yet their coupled formation and excitation mechanisms remain mainly unprobed. The \textit{James Webb Space Telescope (JWST)}, with its high spatial resolution and good spectral resolution, provides unique access to the detection of these molecules.} %leave it empty if necessary  
	{Our goal is to use the first detection of CH$^+$ and CH$_3^+$ infrared rovibrational emission in the Orion Bar and in the protoplanetary disk d203-506 to probe their formation and excitation mechanisms and constrain the physico-chemical conditions of the environment.}
	{We use spectro-imaging acquired using both the NIRSpec and  MIRI-MRS instruments on board JWST to study the infrared CH$^+$ and CH$_3^+$ spatial distribution at very small scales (down to 0.1"), and compare it to excited H$_2$ emission. We study in detail their excitation and, in the case of CH$^+$, we compare the observed line intensities with chemical formation pumping models based on recent quantum dynamical calculations. Throughout this study, we compare the emission of these molecules in two environments: the Bar -a photodissociation region (PDR)- and a protoplanetary disk (d203-506), both irradiated by the Trapezium cluster.}
	{We detect CH$^+$ and CH$_3^+$ vibrationally excited emission both in the Bar and in d203-506. These emissions originate from the same region as highly excited H$_2$ (high rotational and rovibrational levels) and correlates less with the lower rotational levels of H$_2$ ($J'<5$) or the emission of aromatic/aliphatic infrared bands (AIBs). Our comparison between the Bar and d203-506 reveals that both CH$^+$ and CH$_3^+$ excitation and/or formation are highly dependent on gas density. The excitation temperature of the observed CH$^+$ and CH$_3^+$ rovibrational lines is around $T \sim 1500$~K in the Bar and $T \sim 800$~K in d203-506. Moreover, the column densities derived from the rovibrational emission are less than 0.1 \% of the total known (CH$^+$) and expected (CH$_3^+$) column densities. These different results show that CH$^+$ and CH$_3^+$ level populations strongly deviate from local thermodynamical equilibrium (LTE). CH$^+$ rovibrational supra-thermal emission ($v=1$ and $v=2$) can be explained by chemical formation pumping with excited H$_2$ via C$^+$ + H$_2^*$ = CH$^+$ + H. Difference in the population distribution of the H$_2^*$ energy levels between the Orion Bar and d203-506 then result in different excitation temperature. These results support a gas phase formation pathway of CH$^+$  and CH$_3^+$ via successive hydrogen abstraction reactions. However, we do not find any evidence of CH$_2^+$ emission in the JWST spectrum which may be explained by the fact its spectroscopic signatures could be spread in the JWST spectra. Finally, observed CH$^+$ intensities coupled with chemical formation pumping model provide a diagnostic tool to trace the local density. }
	{Line emission from vibrationally excited CH$^+$ and CH$_3^+$ provides us with a new window into the first steps of hydrocarbon gas-phase chemistry in action. This study highlights the need for extended molecular data of detectable molecules in the interstellar medium to analyze the JWST observations.} 
		\maketitle
		%-------------

\section{Introduction}

The methylidyne cation (CH$^+$) and methylcation (CH$_3^+$) are expected to be among the first chemical building blocks of complex organic chemistry in the ultraviolet-irradiated gas \citep{Smith_1992,Herbst_2021}. In dense interstellar clouds, gas-phase ion-neutral reactions are assumed to produce the majority of detected small molecular species. The carbon ion chemistry is initiated by C$^+$, particularly abundant in regions with a high ultraviolet (UV) radiation field, which leads to CH$^+$, CH$_2^+$, and CH$_3^+$ by consecutive hydrogen abstraction \citep[e.g.,][]{Sternberg_1995}. Those small molecular ions are then expected to react with several other species to produce a variety of hydrocarbons. CH$_3^+$ can undergo dissociative recombination producing CH or CH$_2$ which can react again with C$^+$ to produce molecules containing two carbons. Then, the chemistry chain unfolds: unsaturated hydrocarbon ions reacting with small hydrocarbon molecules can result in various long-chain hydrocarbons \citep{Herbst_2021}. In addition, CH$^+$ and CH$_3^+$ are also expected to be at the origin of cyano, amino, and carboxy molecules when reacting with nitrogen- and oxygen-bearing species. 

The emission of these molecules seems particularly enhanced in strongly irradiated environments. Indeed, to be formed, they require either a gas at high kinetic temperature ($T > 400$~K) or a gas with large abundances of rovibrationally excited H$_2$ (see Eq. (\ref{eq:CH3p_form})). In these environments, intense far ultraviolet (FUV, $E < 13.6$ eV) radiation \citep[$G_0 > 10^2$ in Habing units, with $G_0 = 1$ corresponding to a flux integrated between 91.2 and 240 nm of $1.6 \times 10^{-3}$ erg cm$^{-2}$s$^{-1}$,][]{Habing_1968} can produce excited H$_2$ via UV pumping. Thus, these molecules can be used to constrain the physical conditions of these specific regions, providing insight into star and planet formation limited by stellar feedback \citep{Inoguchi_2020}. Moreover, the formation and excitation of CH$^+$ and CH$_3^+$ result from specific physicochemical processes. In addition to constraining physical conditions, they thus can be used to probe UV-driven chemical processes.

CH$^+$ has first been detected in absorption in the diffuse interstellar medium \citep{Douglas_1941}. Since then, CH$^+$ emission has been observed in various environments such as planetary nebulae \citep[e.g.,][]{Cernicharo_1997,Wesson_2010}, massive star-forming regions \citep[e.g.,][]{Falgarone_2010,Bruderer_2010}, disks \citep[e.g,][]{Thi_2011}, and galaxies \citep[e.g,][]{Spinoglio_2012,Rangwala_2014}. In particular, far-infrared (FIR) pure rotational lines of CH$^+$ have been detected in the Orion Bar, a prototypical highly irradiated dense photodissociation region (PDR), with \textit{Herschel}/SPIRE, \textit{Herschel}/PACS and \textit{Herschel}/HIFI \citep{Naylor_2010, Habart_2010,Nagy_2013,Parikka_2017,Joblin_2018,Goicoechea_2019}. In warm gas, it is thought to be the product of the reaction: 
\begin{align}
        \label{eq:CHp_form}
        \ce{C+ + H2($v$^\prime,$J$^\prime) <=> CH+($v$ ,$J$) + H} \qquad \Delta E = +4537 \text{~K},
\end{align}
which is largely endoergic when H$_2$ is in its ground state ($v' = 0$ and $J'=0$). This reaction becomes exoergic when H$_2$ is vibrationally or rotationally excited \citep{Stecher_1972,Jones_1986,Hierl_1997,Agundez_2010,Naylor_2010,Godard_2013,Zanchet_2013,Nagy_2013}. 
While collisional excitation with H and H$_2$ is possible for the lowest rotational levels of CH$^+$, it is unlikely for the rovibrational levels due to their high upper energy levels and high critical densities. 
Moreover, collisional excitation of CH$^+$ is hampered by the high reactivity of this species with both H and H$_2$ \citep{Black_1998}. However, in warm environments, in addition to overcoming the endothermicity of the formation reaction, the internal energy of excited H$_2^*$ can also be used to produce CH$^+$ in an excited state. 
If the lifetime of the molecule is comparable to—or smaller than—the timescale  on which the equilibrium level populations are set, the observed level populations may reflect the initial conditions at formation and chemical formation pumping might be significant. \cite{Godard_2013} show that the pure rotational levels of CH$^+$ with $J \geq$ 2 can be excited by chemical formation pumping during its formation. 

This mechanism has been proposed to explain rotational emission of CH$^+$ in the Orion Bar. This was further confirmed by \cite{Faure_2017} who compared chemical pumping models with observed CH$^+$ line intensities. Their results are  in agreement with \cite{Parikka_2017} (resp. \cite{Goicoechea_2019}) who studied the spatial morphology of CH$^+$ rotational emission lines in the Orion Bar (resp. at very large scales in the Orion Molecular Cloud, OMC-1) at the spatial resolution of \textit{Herschel} (10" or 0.02 pc at 120~$\upmu$m). These authors confirmed a correlation of CH$^+$ with vibrationally excited H$_2^*$. Chemical pumping of the rovibrational levels ($v=1-0$, P(1) - P(10)) of CH$^+$ has been observed in the planetary nebula NGC7027 \citep{Neufeld_2021}. This region has very similar physical conditions to the Orion Bar, with a FUV field intensity around $G_0$ $\sim$ 10$^5$, a gas temperature around $T \sim 1000$~K and a density about $n_{\rm H} \sim 3 \times 10^5$ cm$^{-3}$. 

In contrast to CH$^+$, CH$_3^+$ has only recently been detected in space, outside the solar system, with the James Webb Space Telescope (JWST), in the externally irradiated disk d203-506, near the Orion Bar \citep{Berne_2023,Changala_2023}, and in TW Hya, the nearest T Tauri star with a dusty gas-rich disk \citep{Henning_2024}. Due to its lack of permanent dipole moment, CH$_3^+$ does not have observable rotational transitions and was thus invisible using radioastronomy; it only became detectable thanks to the unprecedented capacities of JWST that enabled the probing of its vibrational spectrum. Because of its only recent detection and spectroscopic complexity, the formation pathway and excitation of CH$_3^+$ remain elusive. It is hypothesized that CH$_3^+$ is formed in the gas phase following successive reactions with H$_2$:
\begin{equation}
\label{eq:CH3p_form}
\ce{C^+ ->[H_2] CH^+ ->[H_2] CH_2^+ ->[H_2] CH_3^+}
\end{equation}

In addition, \cite{Pety_2005} have proposed that the photodestruction of polycyclic aromatic hydrocarbons (PAHs) could be precursors of small hydrocarbons in PDRs. However, \cite{Cuadrado_2015} showed that in highly irradiated PDRs, such as the Orion Bar, the photodestruction of PAHs is not a necessary requirement to explain the observed abundances of small hydrocarbons. Hence, in this region, the gas phase scenario is preferred. In this scenario, CH$_3^+$ would naturally be produced from CH$^+$. Studying CH$_3^+$ in light of CH$^+$ is thus highly relevant. Similar to CH$^+$, in the gas phase, chemical pumping induced by H$_2^*$ could excite CH$_3^+$ as well. 

Chemical pumping is a process expected to excite reactive radicals and molecules in warm irradiated regions. More particularly, hydrides \citep[such as CH$^+$, OH, OH$^+$, HF, ...,][]{Gerin_2016}, produced from reaction with H$_2$, are expected to be particularly sensitive to chemical pumping due to their high reactivity and the enhanced abundance of excited H$_2^*$ in these warm and irradiated regions. Chemical pumping is already employed to explain previous observations of CH$^+$ \citep[e.g.,][]{Godard_2013,Neufeld_2021}, OH$^+$ \citep[e.g.,][]{van_der_tak_2013}, and OH \citep[e.g.,][]{Tabone_2021,Zannese_2024}, and state-to-state reaction rate coefficients have been calculated through ab initio quantum calculation for different reactions in previous studies \citep{Zanchet_2013,Faure_2017,Veselinova_2021,Goicoechea_2022}.
Molecular lines powered by chemical-pumping have also been proven to be a powerful diagnostic of the interstellar medium. \cite{Zannese_2024} showed how the chemical-pumping excitation of OH via O + H$_2$ in its rovibrational levels, detected in the irradiated disk d203-506, can be used to directly derive the formation rate of molecules and the local density from observed intensities.
This method can apply to CH$^+$ and CH$_3^+$ since their formation reactions are similar to OH as it is formed from H$_2^*$.

\begin{figure}
      \centering
      \includegraphics[width=\linewidth]{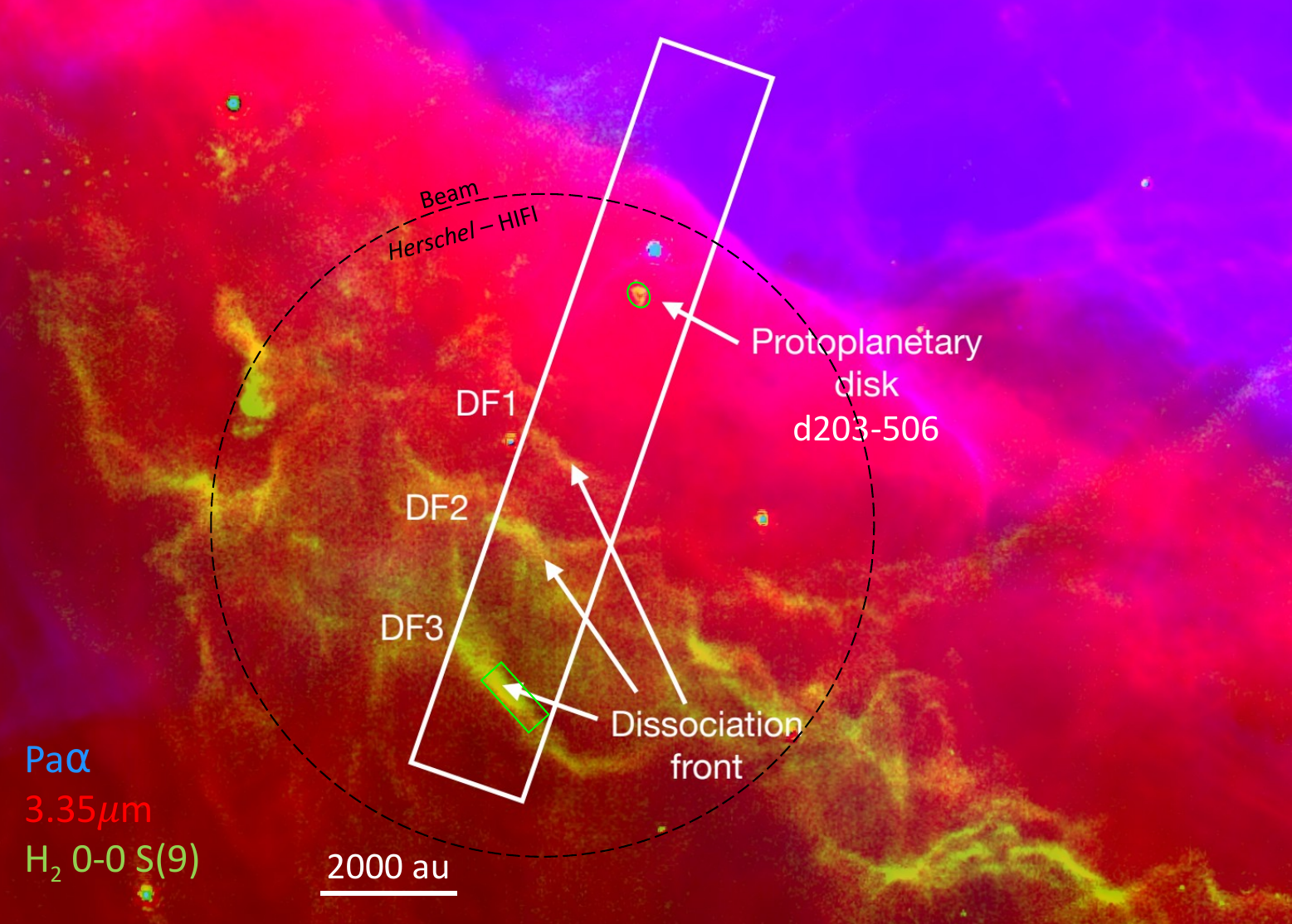}
      \caption{JWST NIRCam composite image of the Orion Bar, located in the Orion molecular cloud \citep{Habart_2024}. Red is the 3.35~$\upmu$m emission (F335M NIRCam filter), blue is the emission of Pa$\alpha$ (F187N filter subtracted by F182M filter) and green is the emission of the H$_2$ 0--0 S(9) line at 4.70~$\upmu$m (F470N filter subtracted by F480M filter). CH$_3^+$ and CH$^+$ emission is detected in the protoplanetary disk and in the dissociation fronts. The white box represents the area where the NIRSpec and MIRI spectroscopy mosaic overlap. The green box corresponds to the apertures used in this paper to derive spectra. We use the aperture from \cite{Peeters_2024} for DF3 and the aperture from \cite{Berne_2024} for d203-506. The dashed black circle corresponds to the largest beam of \textit{Herschel}-HIFI.}
      \label{fig:OB_RGB}
  \end{figure}

In this paper, we present the first detection and analysis of vibrationally excited CH$^+$ and CH$_3^+$ emission in the Orion Bar. This is compared to the detection in the externally irradiated disk (d203-506) located in the line of sight toward the atomic gas rim of the Bar (see Fig. \ref{fig:OB_RGB}). The data were provided by the JWST as part of the program PDRs4All \citep{ERS_2022}. The detection of near-infrared (NIR) and mid-infrared (MIR) lines with the JWST are highly complementary to previous studies as the probed energy levels are very different and the angular resolution ($0.1 - 1$") of JWST is much better than \textit{Herschel}’s (by a factor $10-100$). We are thus able to study the spatial morphology of the emission of these two species. In Sect. \ref{Target}, we present the targets of interest. In Sect. \ref{data_reduc}, we present the observations obtained with JWST/NIRSpec and JWST/MIRI-MRS and the data reduction. In Sect. \ref{line_detection}, we present the detection of CH$^+$ and CH$_3^+$ lines in both the Bar and d203-506. We also study the spatial morphology of their emission and compare it with different tracers in the region such as H$_2$ and the Aromatic/Aliphatic Infrared Bands (AIBs) emission. In Sect. \ref{analysis}, we analyze the excitation of CH$^+$ and CH$_3^+$ to understand the possible chemical routes. We thus study in detail their excitation temperature and compare the difference between the two PDR environments: the edge of a molecular cloud -the Bar and the irradiated surface and photoevaporated wind of an irradiated protoplanetary disk -d203-506. In the rest of the paper, we will qualify the second as the disk to simplify. In this section, we also compare the observations with a simple model including chemical pumping and radiative cascade.
In Sect. \ref{discussion}, we discuss how to use this process as a simple diagnostic to derive local density and formation rate in these regions with this modeling. We also discuss the non-detection of CH$_2^+$ and the other excitation mechanisms that could be at play in these observed transitions.

\section{Targets of interest}
\label{Target}
The Orion Bar, a prototypical highly irradiated photodissociation region \citep[for a review see][]{Hollenbach_1997,Wolfire_2022} located in the Orion Nebula, is the closest site of ongoing massive star-formation \citep[$d=414$ pc,][]{Menten_2007} and acts as a true interstellar laboratory where we can observe the non-thermal processes presented in the introduction. Indeed, this region is exposed to the intense FUV field from the Trapezium cluster, which is dominated by the O7-type star $\theta^1$ Ori C, the most massive star of the Trapezium cluster with an effective temperature $T_{\rm eff} \simeq 40{,}000$~K. The intense FUV radiation field incident on the ionization front (IF) of the Bar is estimated to be $G_0 = 2-7 \times 10^4$ as derived from FUV-pumped IR-fluorescent lines \citep{Peeters_2024}. This intense FUV field shapes the edge of the cloud by ionizing and photodissociating the surrounding gas composing the Orion cloud. This produces the observed PDR, subdivided of several transitions between the HII region, the atomic layer and the molecular region.

Fig. \ref{fig:OB_RGB} shows an RGB view of the Orion Bar observed with NIRCam where both CH$^+$ and CH$_3^+$ emission is detected. The Orion Bar being almost edge-on allows the observation of the transitions between the ionized gas (in blue), the atomic gas (in red) and the H$^0$/H$_2$ transition, or dissociation front (DF, in green). This makes this target particularly useful to study the difference in the observed molecular characteristics, considering the variation of physico-chemical parameters between these zones. The JWST observations \citep{Habart_2024,Peeters_2024} and previous observations with ALMA \citep{Goicoechea_2016} and the \textit{Keck} telescope \citep{Habart_2023} have revealed a complex geometry, where the DFs are filament-like structures at very small scales. In the field of view of MIRI-MRS and NIRSpec, we detect three dissociation fronts with thicknesses around 1". This complexity can be explained by a terrace-field-like structure \citep[see Fig. 5 of][]{Habart_2024}. In the rest of the study, we use the third dissociation front DF3 as a template for all DFs as it is the closest to the surface (and the observer). Indeed, differences in the attenuation of the H$_2$ emission lines indicate that DF1 is located deeper, behind a thicker layer of atomic gas along the line of sight, and DF2 is at an intermediate position between DF1 and DF3 \citep{Habart_2024,Peeters_2024}.

In the field of view of MIRI-MRS and NIRSpec, we also detect a protoplanetary disk (d203-506) located along the line of sight toward the atomic layer. This disk is an almost edge-on disk seen in silhouette against the bright background. The measured radius is $R_{\rm out} = 98 \pm 2$ au and the total mass is estimated to be about 10 times the mass of Jupiter \citep{Berne_2024}. The host star's stellar mass is expected to be below 0.3 M$_{\odot}$ based on kinematic studies with ALMA \citep{Berne_2024}. It is unclear if d203-506 is irradiated by $\theta^1$ Ori C or $\theta^2$ Ori A \citep{Haworth_2023} as the exact location of the disk in the 3D structure is uncertain. However, the intensity of the FUV field at the surface of the disk is expected to be similar to that at the IF as determined with measurements by geometrical considerations (distance between the disk and both stars) and FUV-pumped IR-fluorescent lines \citep[OI, H$_2$, and CI fluorescent lines][]{Berne_2024,Goicoechea_2024}. The observations by the JWST actually reveal the photoevaporative wind surrounding the edge-on disk which is bright in molecular emission (such as H$_2$) rather than the inner disk which is hidden within it.

 \section{Observations and data reduction}
\label{data_reduc}
We used MIRI-MRS and NIRSpec in the integral field unit (IFU) mode observations from the Early Release Science (ERS) program PDRs4All\footnote{\url{https://pdrs4all.org/}, DOI: 10.17909/pg4c-1737}: Radiative feedback from massive stars \citep[ID1288, PIs: Berné, Habart, Peeters,][]{ERS_2022}. Both MIRI-MRS and NIRSpec observations cover a $9 \times 1$ mosaic centered on $\alpha_{\rm J2000} = 05^{\rm h}35^{\rm min}20.4749^{\rm s}$, $\delta_{\rm J2000} = -05^\circ 25$'$10.45$". In this study, we use the full spectro-imaging cubes\footnote{Available on MAST PDRs4All website: \url{https://doi.org/ 10.17909/wqwy-p406}} to study the spatial morphology. The MIRI-MRS data were reduced using version 1.12.5 of the JWST pipeline\footnote{\url{https://jwst-pipeline.readthedocs.io/en/latest/}}, and JWST Calibration Reference Data System\footnote{\url{https://jwst-crds.stsci.edu/}} (CRDS) context jwst\_1154.pmap. In addition to the standard fringe correction step, the stage 2 residual fringe correction was applied as well as a master background subtraction in stage 3 of the reduction. The 12 cubes (4 channels of 3 sub-bands each), all pointing positions combined, were stitched into a single cube  \citep[see][for observation parameters and data reduction details]{Chown_2024,van_de_putte_2024}. The NIRSpec data were reduced using the JWST science pipeline (version 1.10.2.dev26+g8f690fdc) and the context jwst\_1084.pmap of the Calibration References Data System (CRDS) (see \cite{Peeters_2024} for observation parameters and data reduction process). We also use the MIRI-MRS and NIRSpec spectrum, in units of MJy sr$^{-1}$, which were averaged on the apertures given by \cite{Berne_2023} for the disk d203-506 and we used the template spectra\footnote{\url{https://pdrs4all.org/seps/}} from \cite{Peeters_2024}, \cite{Chown_2024} and \cite{van_de_putte_2024} for the Bar.

In this study, the emission of CH$^+$ and H$_2$ can be affected by extinction along the line of sight. The extinction can be neglected in d203-506 as the emission of CH$^+$ and H$_2$ originates from the surface of the wind. However, we need to correct this emission for extinction by the matter in the PDR and the foreground matter. To correct it, we use the $R(V)$-parameterized average Milky Way curve by \cite{Gordon_2023}, evaluated at $R(V) = 5.5$. In DF3, the extinction at the peak of the H$_2$ emission is $A_V = 3.4$, as derived by summing the extinction determined for the foreground ($A_V = 1.4$) and that intrinsic to the atomic PDR ($A_V = 2$)  following \cite{Peeters_2024} and \cite{van_de_putte_2024}. It is important to note that the chosen extinction curve can not attest for the extinction of the H$_2$ 0--0 S(3) \citep[][Sidhu et al. in prep]{van_de_putte_2024}. The evaluation of the proper extinction is difficult in the Orion Bar and will be discussed in Meshaka et al. (in prep). The uncertainty in the correction of extinction has a limited impact on the results of this study, considering other uncertainties.

\section{Detection and spatial distribution of CH$^+$ and CH$_3^+$ rovibrational emission}

\label{line_detection}
		
\subsection{Detection of CH$^+$}

  		\begin{figure}
		    \centering
		    \includegraphics[width=\linewidth]{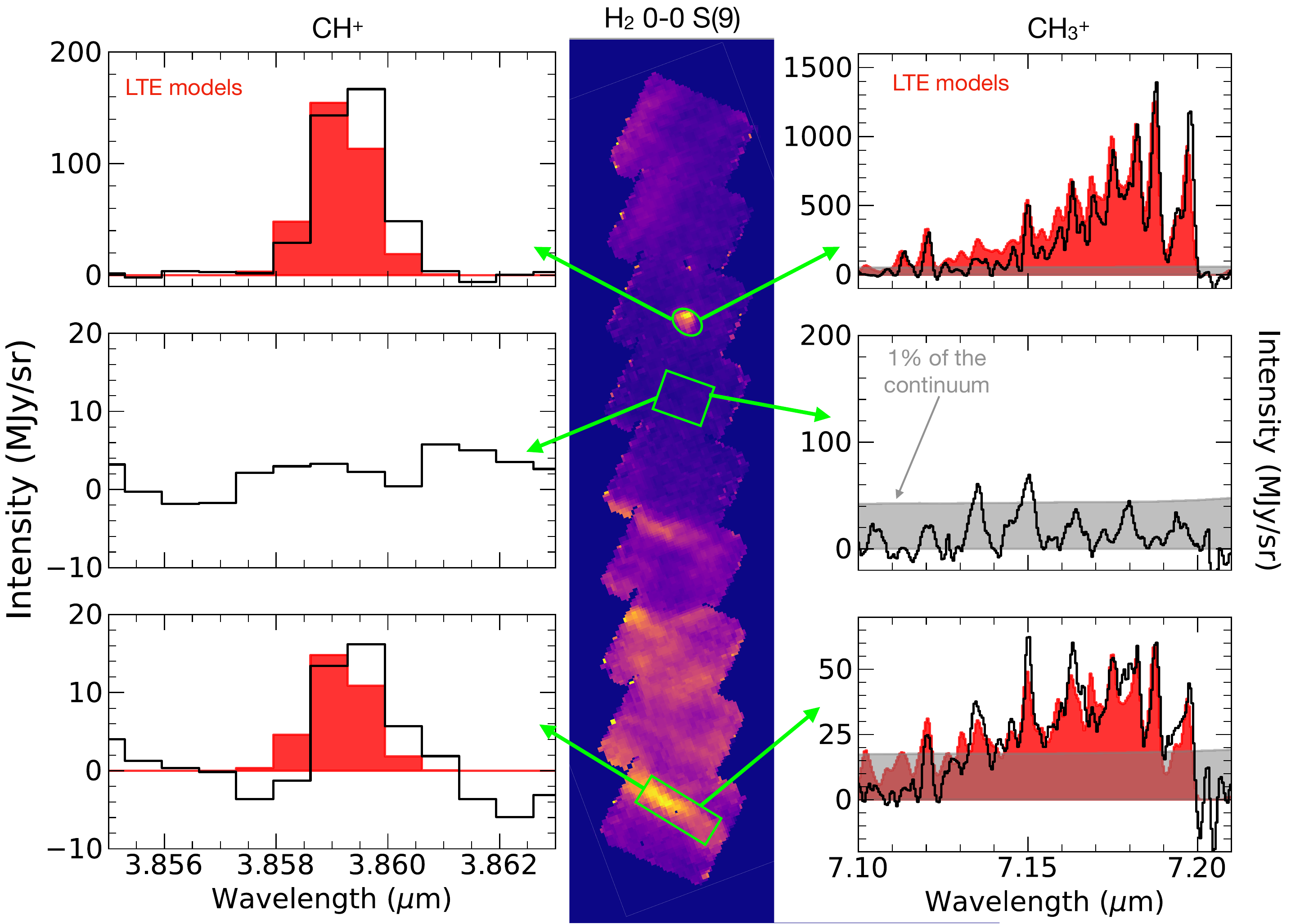}
     
		    \caption{H$_2$ 0--0 S(9) continuum subtracted map at 4.69~$\upmu$m (Sidhu et al. in prep). The green boxes are the aperture used to extract the spectra. (Left) CH$^+$ $v=1-0$ P(5) line at 3.86~$\upmu$m.  (Right) CH$_3^+$ emission from the Q branch around 7.15~$\upmu$m. The solid red graphs are LTE models adapted to the observations (see Sect. \ref{tex} and Fig. \ref{fig:chp_spectra} and \ref{fig:ch3p_spectra}). The gray filling represents 1\% of the continuum. This shows that CH$^+$ and CH$_3^+$ are detected where H$_2$ is bright and not where H$_2$ is faint.}
      \label{fig:template_position}
		\end{figure}

 \begin{figure*}
    \centering
    \includegraphics[width=\linewidth]{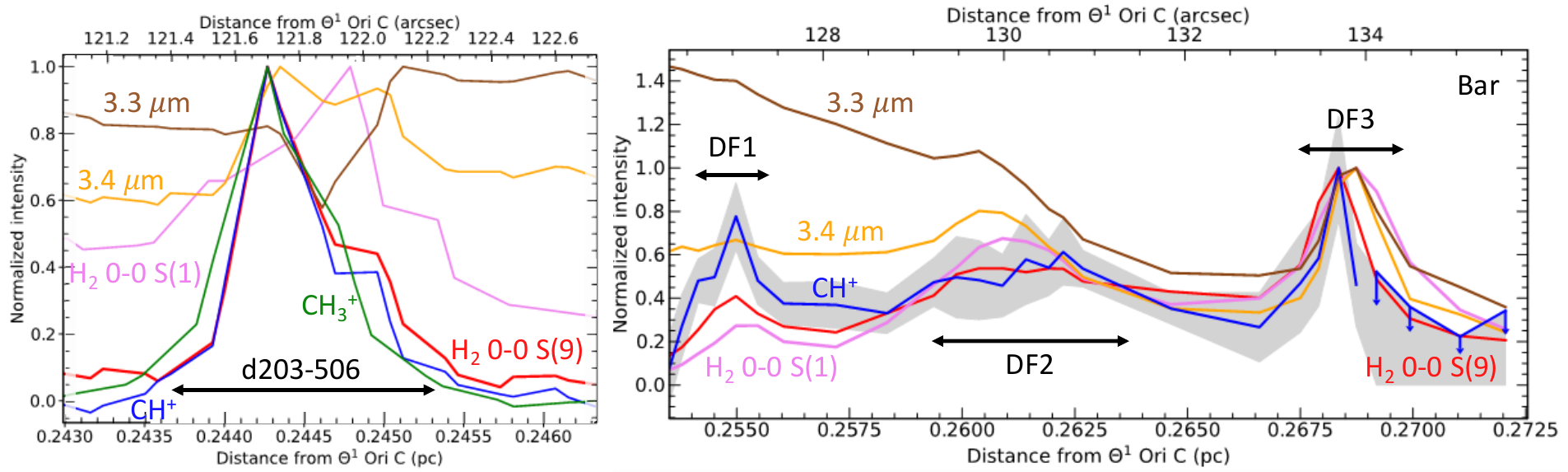}

    \caption{Normalized integrated intensity profiles of the CH$^+$ $v=1-0$ P(5) line, CH$_3^+$ emission between 7.16 and 7.20~$\upmu$m, H$_2$ 0--0 S(1) line, H$_2$ 0--0 S(9) line and AIBs emission at 3.3 and 3.4~$\upmu$m (Left) across the protoplanetary disk d203-506 (Right) across the dissociation fronts in function of the distance to the star $\theta^1$ Ori C. The position of the cut through d203-506 is presented in \cite{Peeters_2024}. In the Bar, each point correspond to the intensity averaged on apertures with width of 2" and height varying from 0.2" to 1.5" to increase the S/N. The maps used to derive the intensity profile of H$_2$ lines (resp. CH$_3^+$ emission) are presented in Sidhu et al. (in prep) (resp. \cite{Berne_2023}). The line 0--0 S(9) was chosen as its wavelength is close to the wavelength range of CH$^+$ infrared rovibrational emission, so the comparison between the line is less affected by extinction.} CH$^+$ and CH$_3^+$ emission follows better the emission of excited H$_2^*$ than the emission of less excited H$_2$ and the emission of AIBs.
    \label{fig:cut}
\end{figure*}
The high sensitivity and good spectral resolution of the JWST allowed the detection of rovibrational emission of CH$^+$ throughout the Orion Bar, in the dissociation fronts and the disk d203-506. Fig. \ref{fig:template_position} shows that CH$^+$ is detected in bright H$_2$ regions (dissociation fronts and d203-506), and is not detected, at the sensitivity of the JWST, in weaker H$_2$ area (atomic region). These rovibrational lines were detected for the first time in NGC7027 with the \textit{NASA’s Infrared Telescope Facility (IRTF)} \citep{Neufeld_2021}. Except for the disk, these lines are very faint (about a hundred times fainter than H$_2$ 1--0 S(1)) so it is necessary to derive spectra from larger apertures than the size of a spaxel to increase the signal-to-noise S/N. 
Line intensities are reported in Table \ref{tab:intensity_chp}. The extinction correction of CH$^+$ emission only increases the line intensities by a factor $\sim$ 1.3. As CH$^+$ is detected with similar intensities in all three dissociation fronts, the intensities reported for the Bar are from DF3. The spectrum of DF3 and d203-506 in the near-infrared is presented in Fig. \ref{fig:chp_spectra}.

We detect CH$^+$ rovibrational lines ($v = 1 \rightarrow 0$ \citep{Peeters_2024} and $v=2 \rightarrow 1$) from 3.5 to 4.4~$\upmu$m with NIRSpec, up to $J=13$ for $v=1$ (upper energy level $E_{\rm up} = 7398$~K) and $J=10$ for $v=2$ ($E_{\rm up} = 9743$
K). The P(5) line at 3.85~$\upmu$m is presented in Fig. \ref{fig:template_position}. CH$^+$ emission from $v=1$ and $v=2$ is split in two branches, the R branch (below 3.62~$\upmu$m) and the P branch (above 3.68~$\upmu$m). The P (resp. R) branch corresponds to rovibrational lines with a change in rotational number $J\rightarrow J+1$ (resp. $J\rightarrow J-1$). Interestingly, we recover the asymmetry between the R branch and the P branch as demonstrated by \cite{Changala_2021}. Due to this asymmetry, the P branch is brighter than the R branch, explaining why we only detect 3 lines from the R branch. For the rest of the study, we thus focus on the P branch.

\subsection{Detection of CH$_3^+$}

CH$_3^+$ has been first detected in the disk d203-506 \citep{Berne_2023,Changala_2023}. The brightest emission of CH$_3^+$,  detected with MIRI-MRS around 7~$\upmu$m, corresponds to its Q branch ($E_{\rm up} \gtrsim 2000$~K) and is presented in Fig. \ref{fig:template_position}.  The observed emission corresponds to the out-of-plane ($\nu_2$) and degenerate in-plane ($\nu_4$) bending modes of the cation \citep{Cunha_2010, Asvany_2018}. The initial assignment has been made possible because the pattern of successive emission lines is characteristic of the spin-statistics of a molecular carrier that would possess three equivalent non-zero-spin atoms (for example, hydrogen atoms). Moreover, the spectrum can be nicely reproduced by models using rotational constants of the order of what is expected from available calculations \citep{KRAEMER_1991,Keceli_2009}. Subsequent high-level quantum calculations have allowed for satisfying reproduction of the observed emission in d203-506, and the spectroscopic assignments were confirmed by experimental measurement of the rotationally-resolved photoelectron spectrum of the $\nu_2$ band of CH$_3^+$ \citep{Changala_2023}.

Here, we present the first detection of this molecule in an interstellar cloud. CH$_3^+$ is detected in the three dissociation fronts as well as in d203-506, where H$_2$ and CH$^+$ emission is bright (see Fig. \ref{fig:template_position}). Unlike CH$^+$, the spectrum of CH$_3^+$ around 7~$\upmu$m is not spectrally resolved. As there are a lot of transitions in this wavelength range, the lines blend together and induce a molecular pseudo-continuum. As is the case for CH$^+$, its emission in the Bar is very faint (about 50 times fainter than H$_2$ 1--0 S(1)) so we also derive the spectra from the same large aperture template. Moreover, the continuum emission at 7~$\upmu$m is strong due to the presence of dust features between 7 and 9~$\upmu$m, which leads to a low line-to-continuum ratio (see Fig. \ref{fig:template_position}). The fringes correction in the MIRI-MRS data reduces them to about 1$\%$ and here the line-to-continuum ratio is about 4$\%$. Hence this complicates the analysis of the CH$_3^+$ feature in the Bar.

\subsection{Spatial distribution}

The spectro-imaging of the JWST allows us to study the spatial morphology of the Orion Bar and compare the emission of different tracers. Fig. \ref{fig:cut} displays the normalized line intensity profile of several lines over the total IFU field of view in d203-506 (Left panel) and in the Bar (Right panel). In the Bar, each point correspond to the intensity averaged on apertures with width of 2" and height varying from 0.2" to 1.5" to increase the S/N. In d203-506, we use the cut presented in \cite{Peeters_2024}. As seen in this figure, the emission of H$_2$ traces the different transition fronts between the atomic layer and the molecular region  (see also Sidhu et al. in prep for in-depth analysis of H$_2$ emission). This emission also traces the photoevaporative wind of the disk d203-506 \citep{Berne_2024}. More precisely, the rotationally excited ($v'=0, J'>5$) and the rovibrational emission of H$_2$ traces the edge of the dissociation front where the temperature and FUV field are higher than further into the molecular cloud, as their excitation is driven by FUV-pumping. The lower rotational levels of H$_2$ ($J'<5$) peak further into the cloud, where the column density is higher and the gas is still warm enough, as they are mainly excited by collisions \citep{van_de_putte_2024}. The spatial separation between the lowest rotational levels of H$_2$ and the highly excited levels is at the limit of the spatial resolution of the MIRI-MRS ($0.3-1$"). The separation between the emission of H$_2$ 0--0 S(9) and H$_2$ 0--0 S(1) is visible in NIRSpec and MIRI-MRS data as highlighted in Fig. \ref{fig:cut} and is about 0.5", which is barely resolved. H$_2$ 0--0 S(1) (in pink) and AIBs emission (in brown) peaks further away than H$_2$ 0--0 S(9) emission (in red) in DF3 (see also \cite{Peeters_2024}). 

We now compare the spatial distribution of the CH$^+$ and CH$_3^+$ emission to that of H$_2$ lines, a very excited one, the 0--0 S(9) line -chosen as its wavelength is close to the CH$^+$ wavelength range so it will be similarly affected by extinction- and a less excited one, the 0--0 S(1) line.  Fig. \ref{fig:template_position} shows a good spatial coincidence between H$_2$ 0--0 S(9) emission and CH$^+$ and CH$^+_3$ emission. The intensity profiles in Fig. \ref{fig:cut} show that CH$^+$ peaks where H$_2$ peaks in both d203-506 and the Bar with more spatial resolution. More precisely, CH$^+$ emission better follows the excited H$_2$ 0--0 S(9) line emission ($E_{\rm up} = 10,261$~K) than the less excited H$_2$ 0--0 S(1) line emission ($E_{\rm up} = 1,015$~K). This comparison between excited H$_2^*$ and CH$^+$ provides a more accurate picture than the previous results \citep{Parikka_2017} which showed that CH$^+$ rotational emission, was at the (lower) spatial resolution of \textit{Herschel}, compatible with a co-spatial emission from H$_2$ 1--0 S(1). These results are in agreement with the necessity of highly excited H$_2^*$ to produce CH$^+$ emission. Moreover, in both panels, it is clear that the CH$^+$ emission profile is closer to the H$_2$ emission than to the AIB emission at 3.3~$\upmu$m and 3.4~$\upmu$m. Overall, this set of result is in agreement with the gas-phase formation (and maybe excitation) of CH$^+$ via reactions between H$_2$ and C$^+$ (see Eq. (\ref{eq:CHp_form})) rather than photodestruction of dust grains.
 
CH$_3^+$ is too faint, or more precisely has too low line-to-continuum ratio, in the Bar to obtain an intensity profile across the mosaic, but the left panel of Fig. \ref{fig:cut} also shows a very good agreement with both CH$^+$ and H$_2$ emission in d203-506 \cite[see also][]{Berne_2023}. Fig. \ref{fig:template_position} also shows CH$_3^+$ to be bright where H$_2$ is bright in the Bar. This good agreement between the spatial distribution of CH$^+$, CH$_3^+$ and H$_2$ is also in favor of a formation of CH$^+$ and CH$_3^+$ in the gas phase.

   \begin{figure}
       \centering
       \includegraphics[width=\linewidth]{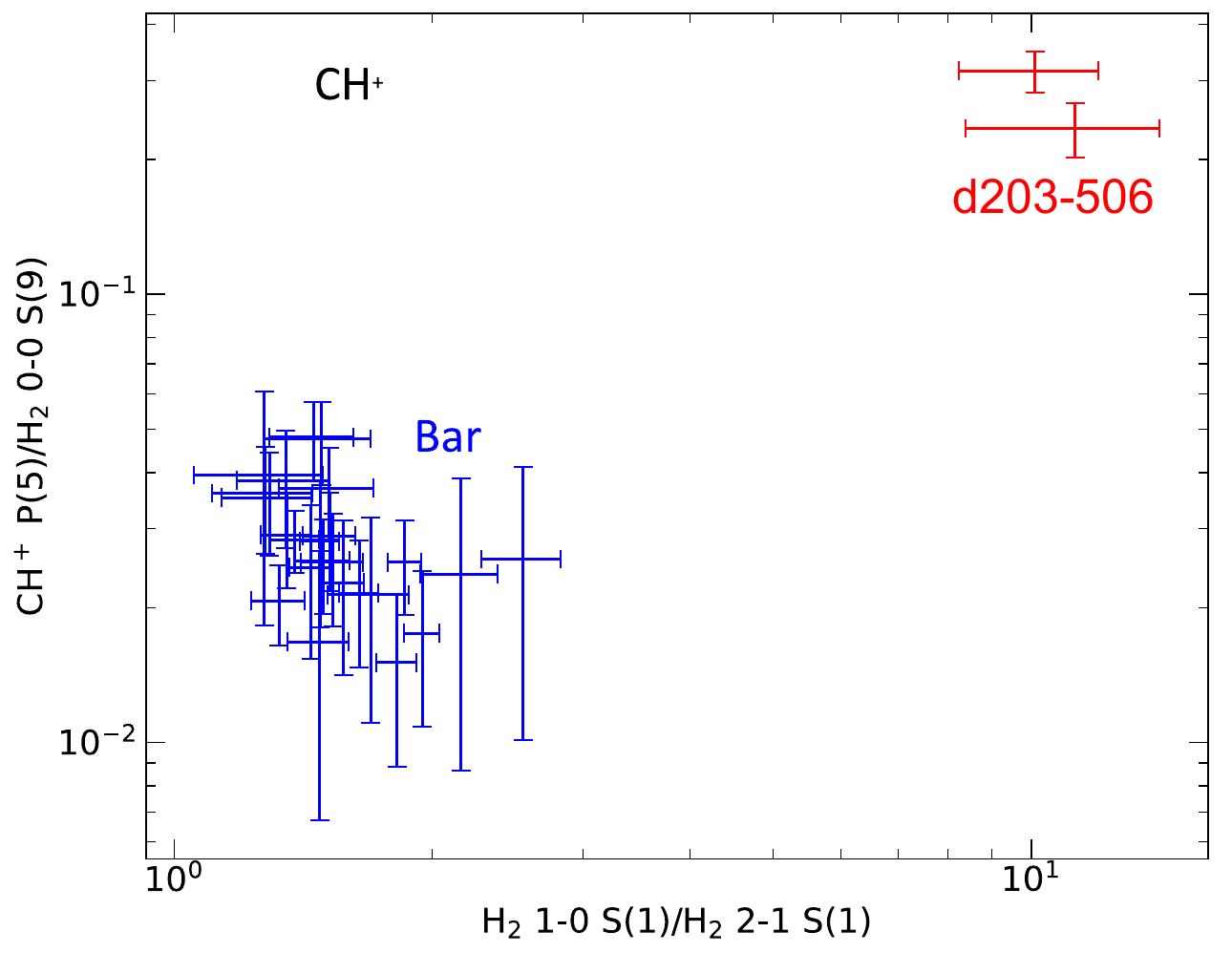}
       \includegraphics[width=\linewidth]{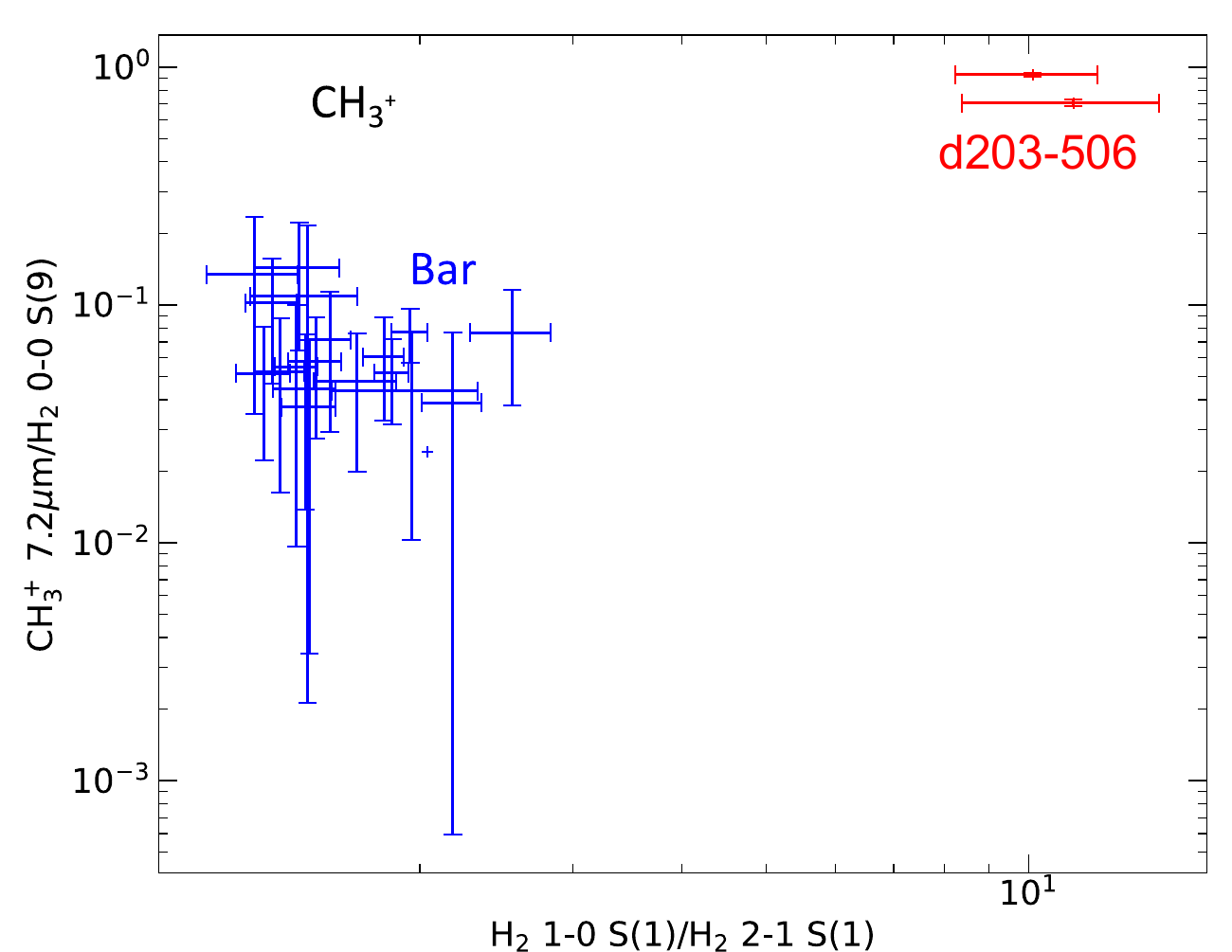}
       \caption{Integrated intensity ratio of CH$^+$ $v=1-0$ P(5) over H$_2$ 0--0 S(9)  (top) and peak intensity ratio of CH$_3^+$ 7.19~$\upmu$m emission over H$_2$ 0--0 S(9) (bottom) as function of the integrated intensity ratio of H$_2$ 1--0 S(1) / H$_2$ 2-1 S(1) which is a tracer of density in these conditions. CH$^+$ and CH$_3^+$ emission seems similarly enhanced at high density in comparison to the emission of H$_2$ 0--0 S(9).}
       \label{fig:line_ratio}
   \end{figure}

\subsection{Dependence on gas density}

While the normalized emission of vibrationally excited CH$^+$ and CH$_3^+$ follows the emission of excited H$_2^*$ well, it is important to note that the intensity ratio of CH$^+$ and H$_2$ 0--0 S(9) for the dissociation fronts and d203-506 is very different.
Fig. \ref{fig:line_ratio} displays the integrated intensity ratio of CH$^+$ P(5) / H$_2$ 0--0 S(9) and peak intensity ratio of CH$_3^+$-7.19~$\upmu$m / H$_2$ 0--0 S(9) as a function of H$_2$ 1--0 S(1) / H$_2$ 2-1 S(1). The H$_2$ 1--0 S(1) / H$_2$ 2-1 S(1) ratio is a tracer of density in dense highly irradiated conditions. In these conditions, collisional excitation of the H$_2$ $v'=1$ $J'=3$ level becomes competitive when the density increases above 10$^5$ cm$^{-3}$. The 1--0 S(1) / 2-1 S(1) line ratio is thus expected to increase from a pure radiative cascade value (about 2) to a collisional excitation value (of the order of 10). This figure shows that for both CH$^+$ and CH$_3^+$, their intensity ratio over H$_2$ 0--0 S(9) is a lot higher in d203-506 than in the Bar (a factor 10 for CH$^+$ and CH$_3^+$). In contrast, the ratio CH$^+$/CH$_3^+$ stays rather constant between the Bar and d203-506, with CH$_3^+$ being slightly more enhanced (by a factor 2) at high density compared to CH$^+$. This suggests that the excitation and/or formation of CH$^+$ and CH$_3^+$ depends more on gas density than the excitation of H$_2$ 0--0 S(9). This is in line with a gas-phase chemical route as density plays a major role in the efficiency of inelastic and reactive collisions.

\section{Analysis of the excitation process}
\label{analysis}

The previous section showing the spatial coincidence between excited H$_2$, CH$^+$, and CH$_3^+$ emission gives insight into the formation pathway of these molecules and favors the gas-phase formation route. However, the spatial distribution is only one piece of the puzzle. In this section, we study in detail the excitation of the species in the Bar and in d203-506, and provide additional support for gas-phase formation and evidence for excitation at formation of at least CH$^+$. 

\subsection{Rovibrational excitation temperature}
\label{tex}
\subsubsection{CH$^+$}

\begin{figure*}
		    \centering

  \includegraphics[width=1\linewidth]{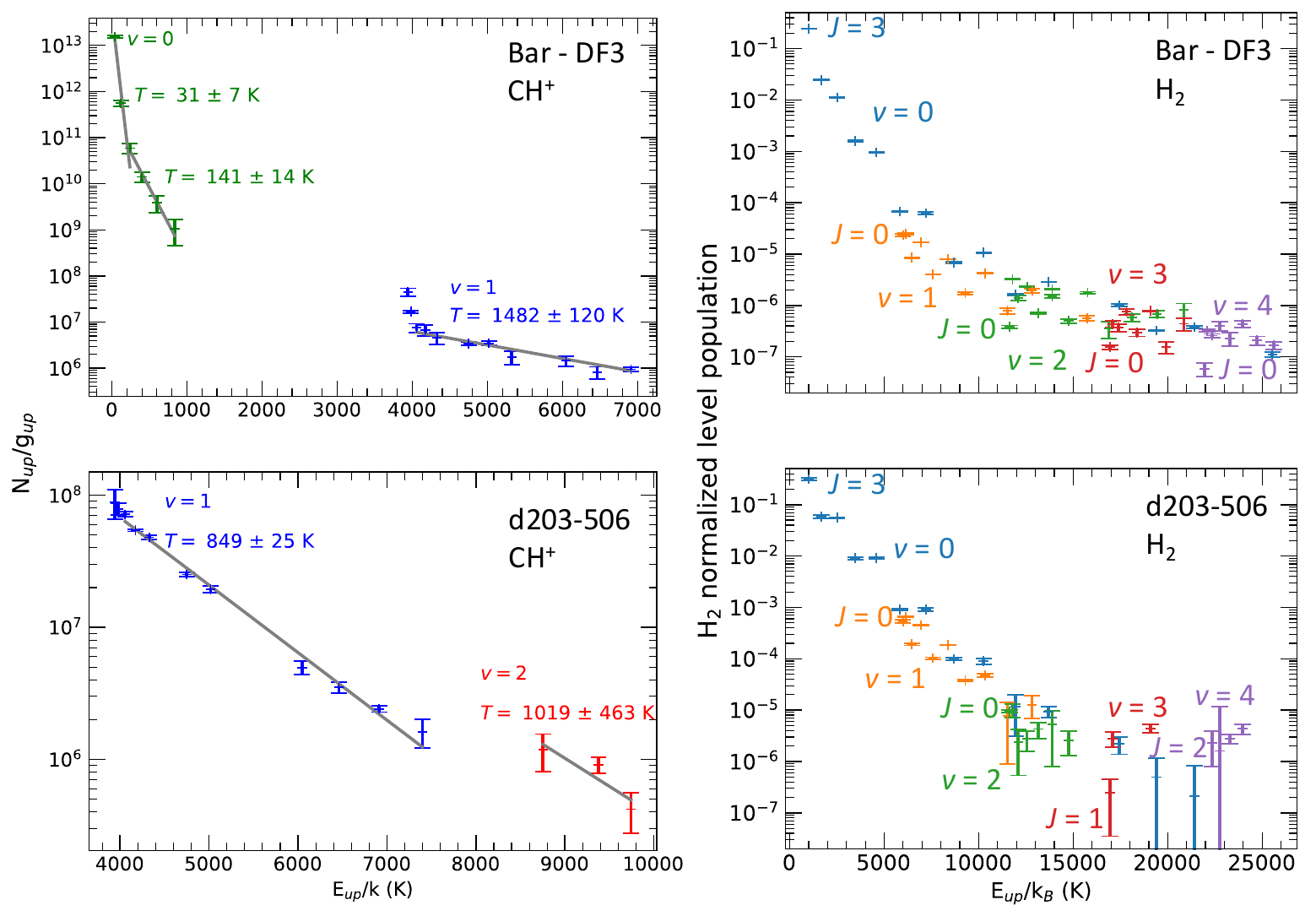}

    \caption{Excitation diagram and level population of CH$^+$ and H$_2$ in d203-506 and in the Bar. (Left) Excitation diagram of CH$^+$ (a) in the Bar and (c) in d203-506. (Right) Level population of H$_2$ normalized to the total column density (see Table \ref{Table:parameters}) (b) in the Bar (d) in d203-506. The difference in the excitation diagram of CH$^+$ can be explained by the difference in H$_2$ population distribution.}
		    \label{fig:diag_rot}
		\end{figure*}

To estimate the excitation temperature of the CH$^+$ rovibrational transitions, we derived the absolute intensities of every line detected with sufficient signal-to-noise (S/N of at least 3) and computed an excitation diagram displayed in Fig. \ref{fig:diag_rot}. This method relies on the assumption that the emission is optically thin which is reasonable because the column density of vibrational CH$^+$ levels are found to be low. To derive the absolute intensities, we fitted the observed lines with a Gaussian coupled with a linear function to take into account the continuum and then integrated the Gaussian function over the wavelengths.

In the Bar, the excitation diagram of the first vibrational mode ($v = 1$) shows that the first two rotational levels are well populated and above $J=2$ the rotational levels align along a straight line (see Fig. \ref{fig:diag_rot}(a)). This means that the excitation of the $v=1$, $J \ge 2$ levels of CH$^+$ follows a Boltzmann distribution, i.e. a single temperature distribution as described by:

\begin{equation}
\ln\left(\frac{N_{\rm up}}{g_{\rm up}}\right)= \ln\left(\frac{N}{Q(T_{\rm ex})}\right)-\frac{E_{\rm up}}{k_{\rm B} T_{\rm ex}},
\label{eq:excit_diagram}
\end{equation}
where $N_{\rm up}$ is the column density of the upper level of the studied transition, $g_{\rm up}$ the upper-level degeneracy,  $N$ the column density, $T_{\rm ex}$ the excitation temperature, $Q(T_{\rm ex})$ the partition function and $k_{\rm B}$ the Boltzmann constant. We note that $N$ corresponds to the total column density of the species only if all the levels follow Eq. (\ref{eq:excit_diagram}). The excitation temperature derived in the Bar from this diagram reaches about 1500~K and reflects a supra-thermal excitation ($T_{\rm ex} > T_{\rm gas}$, i.e. not only a collisional excitation). Indeed, such a high temperature is unlikely to be reached in the Bar at the dissociation front. The gas temperature derived from the H$_2$ lines at the H$^0$/H$_2$ transition is similar and around $T \sim 600$~K \citep[][Sidhu et al. in prep and see Table \ref{Table:parameters}]{van_de_putte_2024}. In addition, models from the Meudon PDR code \citep{Le_Petit_2006} with consistent parameters for the Orion Bar do not predict the temperature to be higher than $T \sim 1000$~K where CH$^+$ and H$_2$ abundances rise \citep[][Meshaka et al. in prep]{Joblin_2018,Zannese_2023,van_de_putte_2024}. 

We also plot the excitation diagram of the pure rotational transitions in the vibrational ground state $v=0$ of CH$^+$ ($40<E_{\rm up} <850$~K) detected in the Bar with \textit{Herschel}/HIFI and \textit{Herschel}/PACS \citep{Parikka_2017,Joblin_2018}. To compare the observations of \textit{Herschel} and JWST, we consider a beam dilution factor as their beam size is different. Following \cite{Joblin_2018}, we considered that in \textit{Herschel} observations, CH$^+$ originates from a 2" wide filament with infinite length, leading to \textit{beam dilution factors} from 0.10 to 0.28 depending on the considered CH$^+$ far-IR lines. Here, we see that the excitation temperature of the $v=0$ levels is significantly lower than the excitation temperature of the $v=1$ levels, except for the $J=0$ and 1 levels within $v=1$. 
The $v=0$ levels are sub-thermally populated due to the high critical densities of the levels and high reactivity of CH$^+$ hampering thermalization of the level via collisions. Overall, the excitation of all levels of CH$^+$ cannot be explained by a single temperature, demonstrating strong deviation to LTE. Another argument for the non-thermal excitation is the observed column density in the first vibrational mode $v=1$ of CH$^+$. In the Bar, we derive a column density $N_{\rm vib}$(CH$^+$) $= (7.6 \pm 2.3) \times 10^{9}$ cm$^{-2}$. The observed column density in $v=1$ is much less than in $v=0$ observed by Herschel \citep[][and see Fig. \ref{fig:diag_rot}]{Joblin_2018}. This is supported by a comparison with predictions of standard PDR models for the Orion Bar physical parameters, which give a total column density for CH$^+$ around $N$(CH$^+$) $\sim$ 10$^{14}$ cm$^{-2}$ \citep{Goicoechea_2025}. This shows that the rovibrational emission of CH$^+$ detected with the JWST does not trace the bulk part of CH$^+$ known to exist (predicted by models and observed by pure rotational lines) but only a very small fraction (lower than 0.1 \%). These results show that the excitation of CH$^+$ in the Bar is not thermalized.

In d203-506, CH$^+$ excitation within a vibrational level ($v=1$ and $v=2$) also follows a Boltzmann distribution with an excitation temperature around $T=850$~K in $v=1$ ($T=1050$~K in $v=2$) which is lower than in the Bar.  This temperature is similar to the gas temperature derived from H$_2$ lines \citep[$T_{\rm gas} \sim 900$~K,][see Table \ref{Table:parameters}]{Berne_2023}. However, the excitation diagram shows an offset between the $v=1$ and $v=2$ levels. Indeed the $v=2$ levels are more strongly populated than expected by extrapolating the line that fits $v=1$ excitation.
From this excitation diagram, we can derive a vibrational temperature between $v=1$ and $v=2$ using the levels $v=1$ $J=7$ and $v=2$ $J=7$ with the equation:
\begin{equation}
    T_{\rm vib} = \frac{E_2-E_1}{k_{\rm B}} / \ln \left(\frac{N_1 g_2}{N_2 g_1}\right). 
    \label{eq:Tvib}
\end{equation}
We measure $T_{\rm vib} \sim 1300$~K which is higher than the rotational temperature $T_{\rm rot} \sim 900$~K. It is also possible that we observe a curvature in the excitation diagram. Indeed, it seems that the high-$J$ levels in the $v=1$ have a higher excitation temperature than the low-$J$ levels. The uncertainties on the data makes it difficult to properly conclude on this matter. However, once again, a unique Boltzmann distribution cannot explain the excitation of all levels of CH$^+$ in d203-506. 
Even though the excitation temperature is close to the gas temperature, it is likely that, overall, levels of CH$^+$ in the disk are also not thermalized.

This analysis shows that the excitation temperature of CH$^+$ is higher and the gas kinetic temperature is lower in the Orion Bar with respect to d203-506. This result indicates that CH$^+$ excitation in the Orion Bar is non-thermal. The different behavior of the excitation temperature highlights the difference in excitation of CH$^+$ in these environments, which can be explained by chemical (formation) pumping. This scenario is explored in the following section.

\subsubsection{CH$_3^+$}

\begin{figure}
        \centering
        \includegraphics[width=\linewidth]{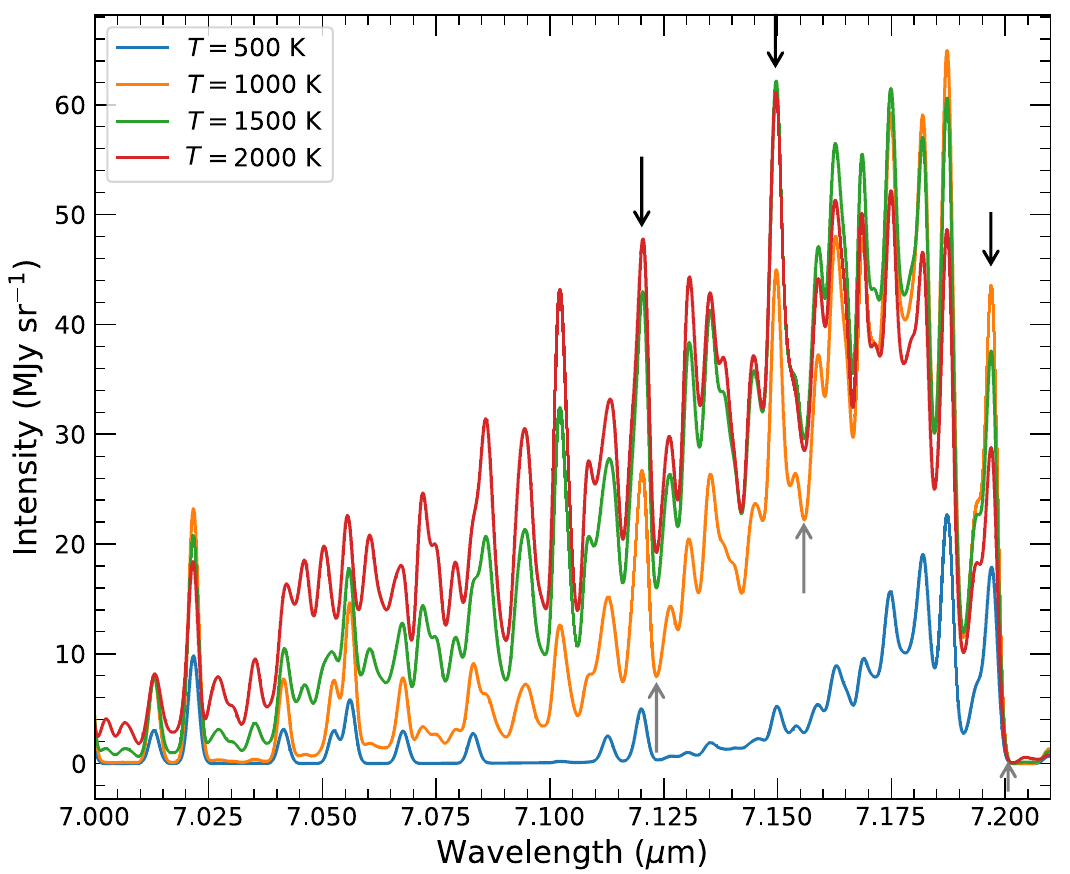}
        \caption{LTE models of the Q branch of CH$_3^+$ at a column density of $N_{\rm vib}$(CH$_3^+$) $= 2 \times 10^{10}$ cm$^{-2}$ at different excitation temperatures using the set of spectroscopic constants from \cite{Changala_2023}. The higher the excitation temperature is, the more intense the lines at shorter wavelengths are. The black (resp. gray) arrows point to the peak (resp. base) of the lines we use for the ratios plotted in Fig. \ref{fig:line_ratio_temperature}.}
        \label{fig:model_ch3+}
    \end{figure}

In order to derive the excitation temperature of CH$_3^+$ from the Q branch only ($\Delta J = 0$), we study the evolution of the shape of the feature as shown in Fig. \ref{fig:model_ch3+}. It appears that the shorter wavelength lines in the Q branch are brighter at higher temperatures, demonstrating that the shape of the Q branch can be used to study the excitation of CH$_3^+$ within the observed vibrational modes.

It is difficult to properly derive the integrated intensities of CH$_3^+$ because the spectrum reveals a broad feature instead of discrete lines and a very weak contrast with the dust continuum emission. Thus, we measure the difference in intensities (in MJy sr$^{-1}$) between the peak and the base of selected prominent and narrow features of the Q-branch that are particularly sensitive to the temperature. 
This method allows us to limit the main uncertainties in this wavelength range: fringes and continuum estimation. Indeed, in the 7~$\upmu$m region, the continuum is very strong due to the rising AIBs between 7 and 9~$\upmu$m which peak at 7.7~$\upmu$m. This makes the slope steep and difficult to fit. Moreover, it is complicated to separate the CH$_3^+$ molecular pseudo-continuum from the dust continuum. Here the line-to-continuum ratio is very low, up to 4$\%$ at best. Hence, the challenge is to determine the dust continuum with a precision that is below 4$\%$ to be able to properly measure the CH$_3^+$ feature. Between the peak and the base of the line, the dust continuum does not drastically vary which makes this measurement almost free of the influence of the continuum estimation. 

\begin{figure}
    \centering
    \includegraphics[width=\linewidth]{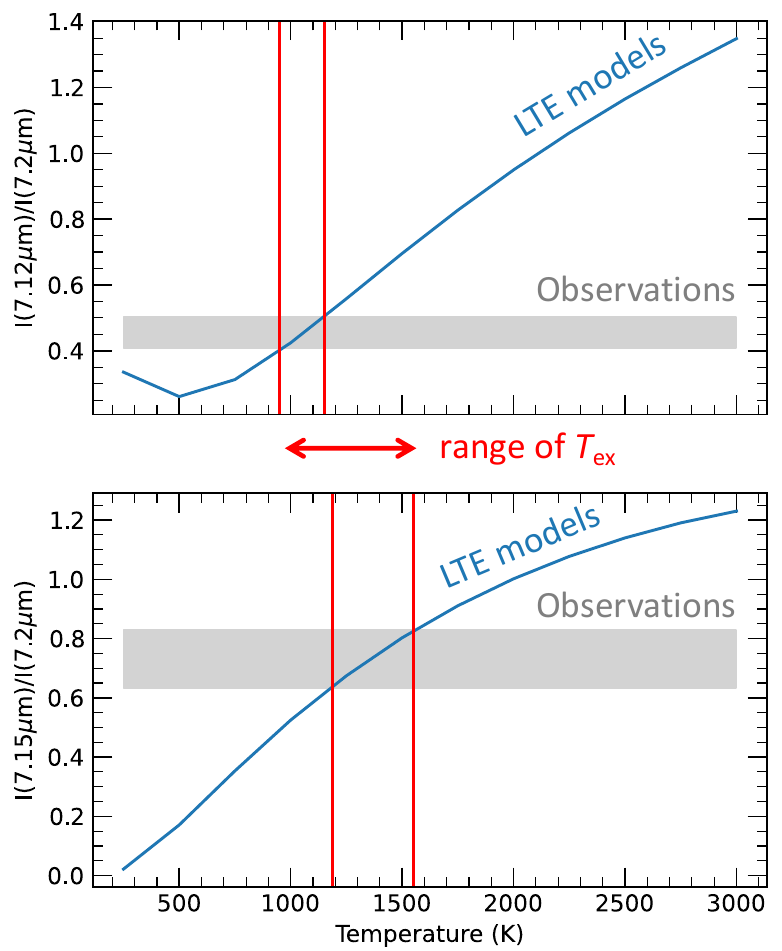}
    \caption{CH$_3^+$ line ratios as a function of temperature. The blue line is the estimation of the ratios from LTE models. The gray area is the measurement of the ratio in the data in DF3. The uncertainties come from the estimation of the continuum. The red lines trace the temperature range where the models cross the observations. (Top) Ratio of the 7.12~$\upmu$m line over the 7.2~$\upmu$m line. (Bottom) Ratio of the 7.15~$\upmu$m line over the 7.2~$\upmu$m line. The excitation temperature of CH$_3^+$ in DF3 is around $1000 - 1500$~K.}
    \label{fig:line_ratio_temperature}
\end{figure}

To estimate the excitation temperature of the CH$_3^+$ emission in DF3, we use the two line ratios 7.12~$\upmu$m/7.2~$\upmu$m and 7.15~$\upmu$m/7.2~$\upmu$m. Fig. \ref{fig:line_ratio_temperature} shows the evolution of these ratios as a function of temperature. The modeled ratios are computed from an LTE model using the spectroscopic data from \cite{Changala_2023}. The measured ratio on the MIRI-MRS spectrum is overlayed in gray. The uncertainties on the measured ratio take into account the variation of the measurement considering the choice of the continuum.  This figure shows that the excitation temperature is between $T=1000$~K and $T=1500$~K in DF3 which is much higher than the gas temperature around 600 K. This excitation temperature is similar to that of CH$^+$. Similarly, using this method, we derive an excitation temperature around $T=700$~K in d203-506 which is in agreement with the temperature derived by \cite{Changala_2023} by fitting a LTE model to the observed spectrum (see Fig. \ref{fig:ch3p_spectra}). As is the case for CH$^+$, the excitation temperature of CH$_3^+$ in DF3 is higher than the gas temperature and higher than the excitation temperature in d203-506. This is also in favor of an excitation by a non-thermal process.

Similarly to CH$^+$, the column density of CH$_3^+$ in the Bar and the d203-506 disk are lower than what is predicted by models when using the rovibrational emission and an LTE assumption. Indeed, we estimate $N_{\rm vib}$(CH$_3^+$) $\sim$ 2 $\times$ 10$^{10}$ cm$^{-2}$ in the Bar and $N_{\rm vib}$(CH$_3^+$)$ = 5 \times 10^{11}$ cm$^{-2}$ in d203-506, whereas the expected column density in the Orion Bar is $N$(CH$_3^+$) $\sim$ 10$^{15}$ cm$^{-2}$ as predicted by models at $n_{\rm H} = 10^5$ cm$^{-2}$ and $G_0 = 10^4$ \citep{Goicoechea_2025}. Similarly to CH$^+$, the observed emission of vibrationally excited CH$_3^+$ accounts for only a very small fraction of the total abundance predicted by models.

\subsection{Evidence for chemical ("formation")-pumping of CH$^+$}

\subsubsection{The model}
\label{chem_pump}
We find the unexpected result that the excitation temperature of both CH$^+$ and CH$_3^+$ is lower in a hotter environment (disk) than in a colder environment (DF3). CH$^+$ excitation due to chemical pumping has already been proposed for its ground vibrational state in the Orion Bar \citep{Godard_2013} and its $v=1$ state in NGC7027 \citep[][]{Neufeld_2021}. In this section, we show that chemical pumping can naturally account for the excitation of CH$^+$ derived from NIRSpec observations using a simple analytical model. We first assume that all the observed lines originate from a single layer and are only excited by chemical pumping by the reaction  C$^+$ + H$_2(v,J)$ $\rightarrow$ CH$^+(v',J')$ + H. To predict line intensities, we also assume that a CH$^+$ cation produced in a given quantum state via C$^+$ + H$_2$ rapidly de-excite by a radiative cascade. We therefore neglect collisions or any other (de)excitation process such as UV or IR pumping. The critical densities of the CH$^+$ rovibrational levels ($n_c \sim 10^{10}$ cm$^{-3}$) are indeed orders of magnitude higher than the expected densities in the Orion Bar or in d203-506 ($n_{\rm H} \sim 10^5-10^7$ cm$^{-3}$). 

The exact population distribution of CH$^+$($v$, $J$) following chemical pumping depends on the state-to-state rate coefficients but also on the local population densities of H$_2$ levels. The probability to form CH$^+$ in a given state $i$ following C$^+$ + H$_2$ is defined as
\begin{equation}
    f_i(\text{CH}^+)_{\text{chem-pump}} = \frac{\sum_j k_{j \rightarrow i}(T) x_j(\text{H}_2)}{\sum_i \sum_j k_{j \rightarrow i}(T) x_j(\text{H}_2)}
\end{equation}
where $x_j({\rm H_2})$ is the level population of H$_2$ normalized to the total population, and $k_{j \rightarrow i}(T)$ is the state-to-state rate coefficient of the reaction ([cm$^{3}$~s$^{-1}$]). We use the state-to-state rate coefficients computed by \cite{Zanchet_2013} with the extension of \cite{Faure_2017}. The available rates are for the reactions from H$_2$ ($v'=1$ $J'=0,1$, $v'=2$ $J'=0$) towards CH$^+$ ($v=0,1,2$, $J$). Hence, we only have state-to-state molecular data for 3 levels of H$_2$ while we have observations of more than 50 levels of H$_2$ in the Orion Bar. In addition, we cannot compare the difference in rates between a vibrational level and a rotational level of H$_2$ with similar energy (e.g., $v'=0$ $J'=8$ and $v'=1$ $J'=0$) as chemical pumping rates coming from highly excited rotational levels of H$_2$ are unavailable. Hence, this study is based solely on the energy of H$_2$ levels, without taking into account the difference between rotation and vibration.
For the other levels of H$_2$ and CH$^+$, we used the extrapolation proposed by \cite{Neufeld_2021} and we normalize it to the rate coefficient of the reaction  C$^+$ + H$_2$ ($v'=1, J'=0$) $\rightarrow$ CH$^+$ ($v=1, J=0$) + H. New quantum calculations are needed to check the validity of this extrapolation. The dependence of these chemical pumping rates on H$_2$ levels and temperature are presented in Figure \ref{fig:proba_chp_h2}.  

The probability $f_i$ to form CH$^+$ in a given state is not to be confused with the probability distribution of CH$^+$ upon which the line intensities depend. The population of a given level is indeed the result of direct production of CH$^+$ in that state and indirect production via a radiative cascade of CH$^+$ produced in higher states. One can then solve the detailed balance equation, including radiative cascade and formation pumping, to compute the column density $N_i$ in a given state and predict line intensities. In this work, we further simplify the detailed balance equation by first noting that the probability to form CH$^+$ in a $v$ state decreases dramatically with the vibrational level $v$. When considering the rotational ladder within a $v$ state, we can therefore neglect the radiative cascade from $v'>v$. We also note that the radiative cascade populating a rovibrational level ($v,J$) is dominated by the ($v \rightarrow v ,J \rightarrow J-1$) transitions. In other words, each vibrational state $v$ can be treated as a separate rotational cascade $J+1 \rightarrow J$ powered by the production of CH$^+$ in the $v$ state with leakage due to the $v \rightarrow v-1$ transitions. Under these simplifying assumptions, the detailed balance equation is
\begin{equation}
    \label{eq:CHp_distrib}
    \left( \sum_m A_{i \rightarrow m}  \right) x_i = A_{i+1 \rightarrow i} x_{i+1} +  \frac{R}{N} \times f_i  
\end{equation}
where $A_{i+1 \rightarrow i}$ is the Einstein coefficient of the pure rotational transition $J+1 \rightarrow J$, $A_{i \rightarrow m}$ is the Einstein coefficient of all the possible transitions from the $i$ level, $R$ is the formation rate of CH$^+$ and N is the total column density of CH$^+$. The left-hand side (LHS) term describes the deexcitation of level $i$ via any downward transitions, essentially pure rotational and rovibrational transitions, the first right-hand side (RHS) term is the population of level $i$ via the radiative cascade, assumed to be dominated by pure rotational transition $J+1 \rightarrow J$, and the second RHS term describes the direct population via formation pumping. This equation is similar to Eq. (C.2) in  \citet{Tabone_2021} with the important difference that Eq. (\ref{eq:CHp_distrib}) includes the $v \rightarrow v-1$ transitions which can be viewed as leakage in the radiative cascade of the rotational ladder.

From Eq. (\ref{eq:CHp_distrib}), one can then compute the intensity (in J cm$^{-2}$ s$^{-1}$ sr$^{-1}$) of a given line as 
\begin{equation}
    \label{eq:intensity_aux}
   I_{ij} = \frac{h  N A_{ij}  \nu_{ij}}{4 \pi} x_i,
\end{equation}
where $ \nu_{ij}$ is the frequency of the considered line. Since Eq. (\ref{eq:CHp_distrib}) is a linear equation in $R/N$, the level population $x_i$ is proportional to $R/N$. Therefore, following \citet{Tabone_2021} and injecting this scaling in Eq. (\ref{eq:intensity_aux}), the line intensity can be rewritten as 
\begin{equation}
   I_{ij} =   \frac{h  \nu_{ij}}{4 \pi} R  \tilde{I}_{ij},
   \label{eq:Iij}
\end{equation}
were $\tilde{I}_{ij} = x_i N A_{ij}/R$ is the normalized line intensity (unitless) which depends only on the nascent distribution $f_i$ and on the Einstein $A$ coefficients. It corresponds to the probability that a CH$^+$ product formed via C$^+$ + H$_2$ transits through the $i\rightarrow j$ transition. Hence, the probability that a CH$^+$ transits through a transition coming from the level $i$ can be written as $\tilde{I}_{i}=\sum_j\tilde{I}_{ij}$. Thanks to this, we can solve Eq. (\ref{eq:CHp_distrib}) by rewriting it as:
\begin{equation}
    \tilde{I}_{i} = \frac{A_{i+1 \rightarrow i}}{\sum_m A_{i+1 \rightarrow m}} \tilde{I}_{i+1} + f_i
    \label{eq:proba_chp}
\end{equation}

To summarize, our simple excitation model uses the distribution of nascent CH$^+$ $f_i$ computed from state-to-rate rate coefficients and state distribution of H$_2$. From this, we compute the distribution of CH$^+$ within each vibrational state using Eq. (\ref{eq:proba_chp}) which is solved iteratively starting from the highest $J$ level of the considered vibrational state for which $x_{i} \simeq 0$.

\subsubsection{Application}

Thanks to both MIRI-MRS and NIRSpec observations, the population densities of H$_2$ are directy measured both in the Bar and in d203-506. The H$_2$ level population diagrams are plotted in Fig. \ref{fig:diag_rot}.

To take into account H$_2$ levels which are not observed with JWST but are significantly populated by FUV-pumping \citep[detected up to $v'=12$ with IGRINS,][]{Kaplan_2017,Kaplan_2021}, we consider that they are populated following a Boltzmann distribution at the gas temperature as a lower limit. For the upper limit, we consider that all levels of H$_2$ with upper energy levels $E_{\rm up}$ < 30,000~K are as populated (considering degeneracy) as the last observed level of the same vibrational mode. Beyond 30,000~K, the population of the levels are at least 7 orders of magnitude lower than the low pure rotational levels whereas the state-to-state coefficients reach a constant value which is only higher by 5 orders of magnitude, so they can be neglected. The gas temperature is also estimated in both environments thanks to the pure rotational lines of H$_2$ \citep[Sidhu et al. in prep]{van_de_putte_2024,Berne_2023}. 

Here, we assume that the excitation of CH$^+$ is driven by chemical-pumping, and that there is a negligible impact of radiative pumping and inelastic collisions. We take into account only the first four vibrational modes of CH$^+$. Using the state-to-state coefficients, the distribution of H$_2$ observed with JWST and extrapolated, and the derived gas temperature, we can calculate the probability that CH$^+$ eventually de-excites through observed transitions  following its excitation by chemical-pumping.

\begin{figure}
    \centering
    \includegraphics[width=\linewidth]{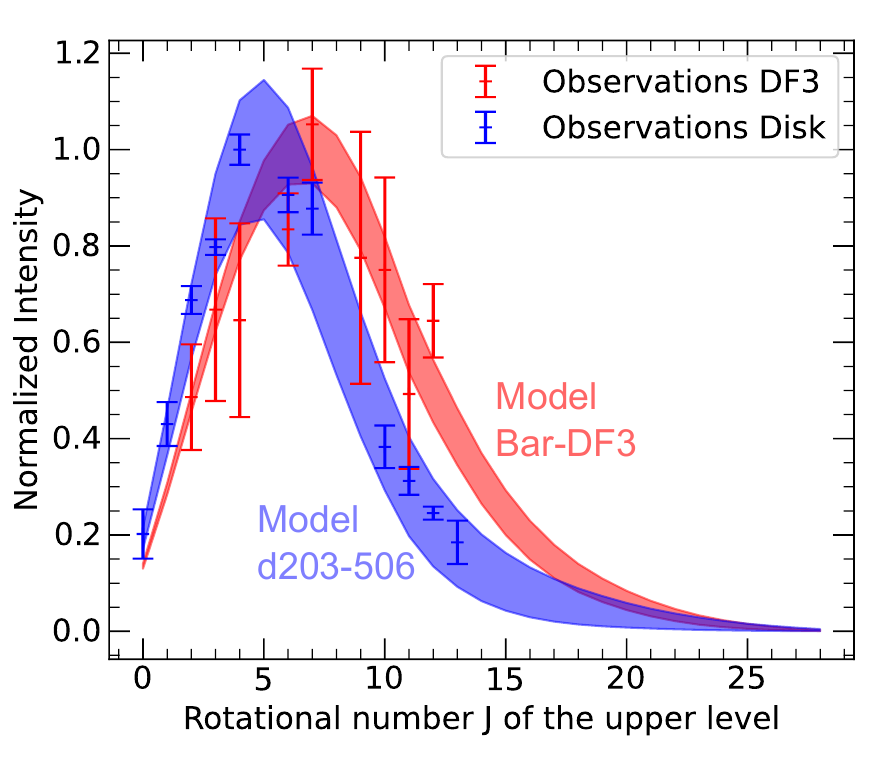}
    \caption{Normalized intensities $\Tilde{I}_{ij}$ of CH$^+$ rovibrational transitions $v=1\rightarrow 0$, $J \rightarrow J+1$ following chemical-pumping considering observed distribution of H$_2$ and temperature. The red (resp. blue) line corresponds to the intensities of CH$^+$ considering the temperature and H$_2$ population densities in the Bar (resp. in d203-506). The shaded areas indicate the range between the upper and lower limits as determined in Sect. \ref{chem_pump}. Red (resp. blue) crosses correspond to the normalized intensities of CH$^+$ transitions observed in the Bar (resp. disk).}
    \label{fig:distrib_chp}
\end{figure}

Fig. \ref{fig:distrib_chp} displays the calculated normalized intensities $\Tilde{I}_{ij}$ of the rovibrational transitions $v=1\rightarrow 0$, $J \rightarrow J+1$ of CH$^+$ considering both direct and indirect chemical pumping (see Eq. (\ref{eq:proba_chp})) in the Bar and in d203-506. The chemical-pumping model shows a good agreement with observations, except for the levels $v=1$ $J=0,1$ in the Bar, which could be excited by other processes (see Sect. \ref{discussion}). Excluding these first two levels, we find that chemical pumping naturally reproduces the difference in excitation temperature in d203-506 and in the Bar. Indeed, the figure shows a maximum in the intensities of CH$^+$ for upper level rotational numbers between $J=5-10$ in the Bar versus $J=2-7$ in the disk. This explains the higher excitation temperature detected in the Bar compared to the disk. This difference in rotational excitation reflects the difference in H$_2$ level population densities in these environments. In d203-506, the gas temperature is higher than in the Bar, hence the excited pure rotational lines ($v'=0$ $J'=5-10$) are more populated proportionally to other levels in the disk than in the Bar. Excitation of CH$^+$ in the disk is thus dominated by formation by C$^+$ reacting with the pure rotational levels of H$_2$ which roughly follow LTE ($J'<9$). In the Bar, those levels are much less populated and therefore contribute less to the formation of CH$^+$. Indeed, H$_2$ $J'<9$ accounts for about 18\% of the excitation of CH$^+$ in $v=1$ in the disk against less than 2\% in the Bar. If the pure rotational levels excited by collisions are more populated, they have however lower energies than FUV-pumped rotational and rovibrational levels $v'=0$ $J'>8$ and $v'>0$. They will dominate the excitation of CH$^+$ but they will mostly be able to excite lower levels such as $v=1$ $J=2-7$. The peak of the intensities being around these levels explains the rather "low" excitation temperature of CH$^+$ in the disk compared to the Bar. On the contrary, in the Bar, the excitation of CH$^+$ is expected to be dominated by C$^+$ reacting primarily with FUV-pumped H$_2$ levels which have higher energies and therefore populate higher levels of CH$^+$, such as $J=5-10$. Thus, the excitation temperature of CH$^+$ is higher compared to the disk. This result suggests that CH$^+$ is excited by thermally excited levels of H$_2$ in regions with high gas temperature and by FUV-pumped levels of H$_2$ in highly irradiated regions with lower temperature. Figure \ref{fig:distrib_comp_ETL} shows the importance of FUV-pumped levels of H$_2$ in the excitation of CH$^+$ ($v=1$) in DF3 when thermalized levels of H$_2$ can almost explain the excitation of CH$^+$ ($v=1$) in d203-506. These results are mostly based on extrapolated rates, which do not take into account the difference between H$_2$ vibrational and rotational excitation. Additional quantum calculations are necessary to confirm rotational levels of H$_2$ are truly sufficient to excite CH$^+$ in d203-506.

\begin{figure}
    \centering
    \includegraphics[width=\linewidth]{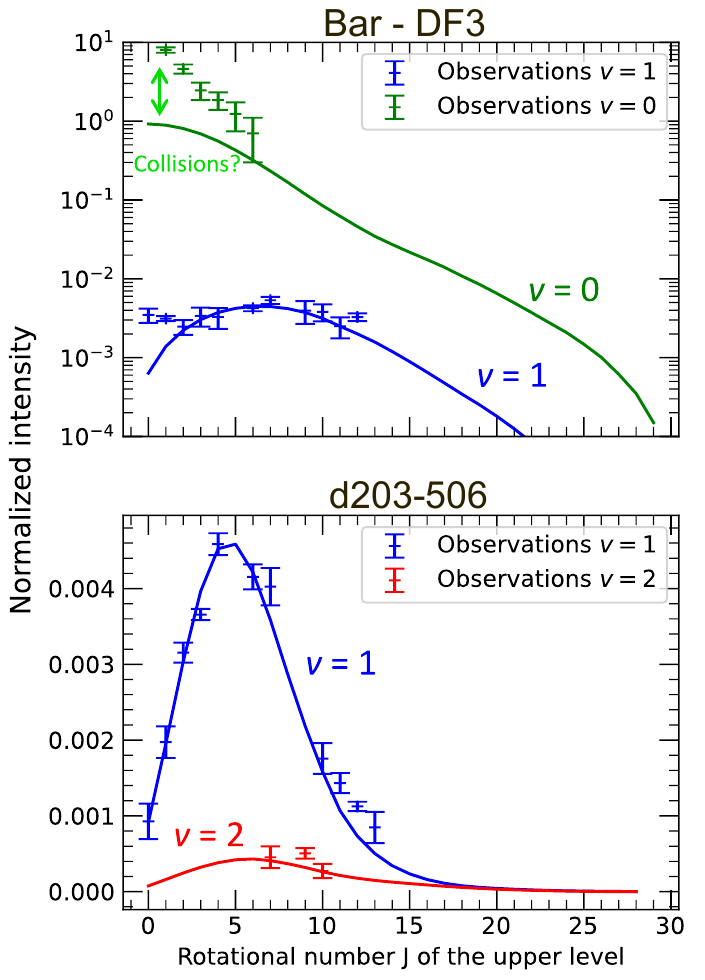}
    \caption{Normalized intensities $\Tilde{I}_{ij}$ of CH$^+$ of $v=0 \rightarrow 0$, $J \rightarrow J-1$ (green), $v=1 \rightarrow 0$, $J \rightarrow J+1$ (blue), $v=2 \rightarrow 1$, $J \rightarrow J+1$ (red) following chemical-pumping considering the observed population densities of H$_2$ and temperature in the Bar and in d203-506. Emissions from $v=1$ and $v=2$ levels are observed with NIRSpec. Emission from $v=0$ levels are observed with \textit{Herschel}/PACS and \textit{Herschel}/HIFI and we considered that the emission comes from a 2" wide filament \citep{Joblin_2018}.}
    \label{fig:distrib_v0v1v2}
\end{figure}

In addition to reproducing the emission from the $v=1$ level well, Fig. \ref{fig:distrib_v0v1v2} shows that the chemical pumping model also accounts for the emission from the $v=2$ levels detected in NIRSpec in the disk. The observed ratio between $v=1$ and $v=2$ is particularly well reproduced.
We can see that chemical pumping is also compatible with the $v=0$ emission detected in the Bar with \textit{Herschel}/HIFI and \textit{Herschel}/PACS \citep{Parikka_2017,Joblin_2018}. Indeed, chemical pumping model which reproduces $v=1$ and $v=2$, does not overestimate the emission expected in $v=0$. First, the fact that most CH$^+$ line intensities of $v=0$ are too high may be explained by the uncertainty on the \textit{beam dilution factor} which can be underestimating the physical size of the CH$^+$ emitting region. The over-population of the lower-$J$ levels and the fact that the population of the $v=0$ $J=6$ level is compatible with chemical pumping taking into account uncertainties is also highlighting the importance of inelastic collisions in the excitation of the low-$J$ levels. This result is in agreement with \cite{Godard_2013} who have shown that, in the Orion Bar, high-$J$ transitions are mostly driven by chemical pumping, but lower-$J$ lines are affected by inelastic collisions. Indeed in Table 4 of \cite{Godard_2013}, they show that collisions account for as much as 80\% of the excitation of $J=1$ and 40\% of the excitation of $J=6$.

\section{Discussion}
\label{discussion}
\subsection{Diagnostics on chemistry and physical conditions}
In the previous section, we used the relative line intensities to support that the vibrational bands of CH$^+$ detected in the Orion Bar and in d203-506 are excited by chemical pumping. When a line is excited by chemical pumping, its absolute intensity carries crucial information about the local conditions.

\subsubsection{Formation rate of CH$^+$ by C$^+$ + H$_2^*$}

 Following the formalism of \cite{Zannese_2024} used for OH chemical-pumping excitation, we can estimate a formation rate of CH$^+$ from the intensities measured in the observations and the model of chemical-pumping. Here we assume that, in both environments, the impact of radiative pumping and inelastic collisions is negligible. Furthermore, the rovibrational excited lines are found to be optically thin. 
Hence, the intensities $I_{ij}$ are simply proportional to $R$, the formation rate of CH$^+$ as described in Eq. (\ref{eq:Iij}). Thus, the formation rate can be calculated as follows:

\begin{equation}
    R = \frac{4\pi I_{ij}}{h \nu_{ij} \Tilde{I_{ij}}}
\end{equation}

From the observed intensities and the population distribution of CH$^+$, we derive $R=(2.0-6.0)\times10^{10}$ cm$^{-2}$ s$^{-1}$ in the Bar and $R=(1.5-4.5)\times10^{11}$ cm$^{-2}$ s$^{-1}$ in d203-506. The formation rate of CH$^+$ is about 7 times higher in the disk than in the Bar. A brief summary of the important results of our study is presented in Table \ref{Table:parameters}.

\subsubsection{Gas density}

\cite{Zannese_2024} showed that chemical pumping models of O + H$_2$ can be used to estimate the local density. Here, we follow this formalism because of the similarity of processes (O + H$_2$ vs C$^+$ + H$_2$). Hence, the formation rate, integrated over the line of sight, of CH$^+$ via C$^+$ + H$_2$ corresponds to

\begin{equation}
        R \equiv \int_{z} \sum_i \sum_j  k_{j \rightarrow i}(T) x_j({\rm H_2}) n_{\rm H_2} n_{\rm C^+} \text{d}z = \int_{z} k(T,x_j({\rm H_2}))  n_{\rm H_2} n_{\rm C^+} \text{d}z.
        \label{eq:F}
\end{equation}
where $n_{\rm C^+}$ and $n_{\rm H_2}$ are the number densities of ionized carbon and molecular hydrogen and $k(T,x_j({\rm H_2}))$ is the total formation rate ([cm$^{3}$~s$^{-1}$]): 
\begin{equation}
        k \equiv  \sum_i \sum_j  k_{j \rightarrow i}(T) x_j({\rm H_2}).
        \label{eq:k_tot}
\end{equation}

Assuming a homogeneous medium, the local density can be estimated from the inferred value of $R$ from Eq. (\ref{eq:F}) as:
\begin{equation}
\label{eq:nH}
   n_{\rm H} = \frac{R}{k N({\rm H_2}) x({\rm C^+})},
\end{equation}
where $N(\rm H_2)$ is the column density of H$_2$, $x(\rm C^+)$ is the ionized carbon abundance, and $n_{\rm H}$ is the total number density of hydrogen nuclei. 

In both environments, the ionized carbon abundance can be taken around $x($C$^+$) $\simeq 1.4 \times 10^{-4}$ \citep{Sofia_1997}. In d203-506, we measure a column density of warm H$_2$ of $N(\rm H_2) = 8.7 \times 10^{19}~$cm$^{-2}$ and a temperature of $T \simeq 850$~K \citep{Berne_2023}. We derive a total rate coefficient of $k=(2.5-
2.6) \times 10^{-12}$ cm$^3$ s$^{-1}$ from Eq. (\ref{eq:k_tot}), and using the H$_2$ excitation diagram and the state-specific rates of \cite{Zanchet_2013,Faure_2017,Neufeld_2021}. Using the estimate of $R$ = $(1.5-4.5) \times 10^{11}$ cm$^{-2}$ s$^{-1}$ from the CH$^+$ near-IR lines, we derive $n_{\rm H}$ = $(0.6-2.0) \times 10^7$ cm$^{-3}$. Interestingly, the intensity derived by this diagnostic gives a similar value as the very similar diagnostic made by \cite{Zannese_2024} with chemical pumping for the OH molecule via O + H$_2$. This is also in agreement with the estimation made from a different approach by \cite{Berne_2024} using H$_2$ lines and the Meudon PDR Code \citep{Le_Petit_2006}. 

Similarly, in the Bar, the column density of warm H$_2$ is $N(\rm H_2) = 1.6 \times 10^{21}~$cm$^{-2}$ and the temperature is $T \simeq 570$~K from the rotational lines \citep[][Sidhu et al in prep]{van_de_putte_2024}. From the estimation of temperature and excitation of H$_2$, we derive the coefficient rate $k= (2.1-2.2) \times 10^{-13}$ cm$^3$ s$^{-1}$. Using the estimate of $R = (2.0-6.0 )\times10^{10}$ cm$^{-1}$ s$^{-1}$ from the CH$^+$ near-IR lines, we derive $n_{\rm H} = (0.6-1.5)\times10^6$ cm$^{-3}$. This value leads to a thermal pressure for the dissociation front to be around $P_{\rm gas} \simeq 3-7\times 10^8$~K cm$^{-3}$. One should know that the uncertainties of extrapolated rates are difficult to estimate. However, we assume that they are negligible considering other sources of uncertainties that we discuss below.

This density estimate is higher than what is found from the H$_2$ emission \citep[$P_{\rm gas} \sim 5\times 10^7 - 2\times 10^8$~K cm$^{-3}$,][Meshaka et al. in prep, Sidhu et al. in prep]{Joblin_2018,van_de_putte_2024}. Whether the thermal pressure is constant all over the dissociation front \citep[e.g.,][]{Bron_2018} is difficult to affirm as the spatial resolution is not sufficient to actually probe the variation of density and temperature across the DF. Here, we assume a single-layer model which does not consider any variation in physical conditions. 
We have shown previously that the CH$^+$ lines are more sensitive to the density than the highly excited H$_2$ (see Fig. \ref{fig:line_ratio}). They may be more influenced by the higher density part of the front and thus trace a higher thermal pressure than what is derived from the overall emission of H$_2$ (rotational and rovibrational). 
Moreover, to increase the S/N of the CH$^+$ line detection, we derived the intensities from a large aperture. This means that the population densities of H$_2$ is also averaged. By averaging the population densities of H$_2$, we probably underestimate the proportion of H$_2$ in highly excited states in comparison to the lowest excited states at the position where CH$^+$ peaks. Underestimating these levels leads to overestimating the gas density. Indeed, as they have more internal energy, they are more reactive and will enhance the formation of CH$^+$ in excited states.
We can use models to estimate how the density can be impacted by the averaging of H$_2$ population densities. We use the best-fit model from the Meudon PDR Code \citep{Le_Petit_2006} reproducing H$_2$ emission at $P_{\rm gas}$ = 5 $\times$ 10$^7$~K cm$^{-3}$ (Meshaka et al. in prep). We calculated H$_2$ population densities, H$_2$ column density and temperature at the peak of the line H$_2$ 1--0 S(1) and the peak of the line H$_2$ 0--0 S(1). The value of $k \times N$(H$_2$) averaged around the peak of H$_2$ 1--0 S(1) is about 5 times larger than around the peak of H$_2$ 0--0 S(1). Interestingly, this factor is similar even in a higher thermal pressure model ($P_{\rm gas} = 5 \times 10^8$~K cm$^{-3}$). Thus, using the equation Eq. (\ref{eq:nH}), the derived density can be overestimated by as much as a factor 5 according to models. The use of self-consistent models, such as the Meudon PDR Code, including CH$^+$ chemical pumping, is necessary to provide more accurate predictions considering the steep variation of physical parameters. It will also be useful to validate our zero-dimensional models by comparing them to PDR models, taking into account the depth of the cloud.  However, this further modeling is beyond the scope of this paper.

\begin{table*}[!h]
  \begin{center}
  \caption{Parameters derived from the analysis of the CH$^+$ and H$_2$ lines}
  
\label{Table:parameters}
\begin{tabular}{|l|l|c|c|}
   \hline
Diagnostics & Quantity & Measured value in the Bar & Measured value in d203-506 \\[0.5ex]
   \hline
   \hline
 H$_2$ lines  &$N$(H$_2$)\tablefootmark{a} &  $(1.6 \pm 0.1) \times 10^{21}$ cm$^{-2}$ &  $(8.7 \pm 0.8) \times 10^{19}$ cm$^{-2}$ \\[0.5ex]
&$T$\tablefootmark{a}  &  566 $\pm$ 7~K &  836 $\pm$ 13~K \\[0.5ex]
 \hline
CH$^+$ near-IR lines  & $F$ (rate  C$^+$+H$_2$)\tablefootmark{b,c}       &  $2.5-5.1 \times 10^{10}$ cm$^{-2}$ s$^{-1}$  &  $1.5-5.3 \times 10^{11}$ cm$^{-2}$ s$^{-1}$   \\[0.5ex]

 & $n_{\rm H}$\tablefootmark{c} &   $0.5-1.1 \times 10^6$ cm$^{-3}$       &   $0.5-1.4 \times 10^7$ cm$^{-3}$      \\[0.5ex]
   \hline
\end{tabular}
\end{center}
\tablefoot{\tablefoottext{a}{The uncertainties of these values do not take into account calibration effects. They could be increased by a factor of $2-3$ if the uncertainty associated with calibration effects is equal to 20\%.}
\tablefoottext{b}{Number of CH$^+$ molecules formed per unit area and time via either formation route (in cm$^{-2}$ s$^{-1}$).}
\tablefoottext{c}{The inferred value would be reduced if inelastic collisional excitation were to play a significant role.}}
\end{table*}

\subsection{Alternative excitation processes}

To ensure that the observed lines are only excited by chemical pumping, we need to verify that other excitation processes cannot explain the observed spectrum. Moreover, the first two lines of CH$^+$, $v=1$ P(1) and P(2), detected in the Bar cannot be explained by the chemical pumping model (see Fig. \ref{fig:distrib_v0v1v2}).

\subsubsection{IR pumping}

IR pumping can be an efficient process to excite the first vibrational state since it takes the absorption of only one near-infrared photon to bring CH$^+$ to a $v=1$ $J$ state.
If IR pumping is relevant to excite the observed levels, then the brightness temperature of the infrared radiation $T_{\rm IR}$ around $3.5-4.5$~$\upmu$m has to be similar to the vibrational excitation temperature $T_{\rm vib}$,

\begin{equation}
    T_{\rm IR} = \frac{h c}{\lambda k_{\rm B}}/\ln{\left(1+\frac{2 h c}{J_\lambda \lambda^3}\right)}
\end{equation}

where $J_\lambda$ is the intensity of the infrared radiation field at the wavelength $\lambda$ (in W m$^{-2}$ Hz$^{-1}$ sr$^{-1}$). If the field is isotropic, $J_\lambda = I_\lambda$ where $I_\lambda$ is the specific intensity observed in the line of sight, corrected for extinction at the wavelength $\lambda$. Around $3.5-4.5$~$\upmu$m, $I_\lambda \sim 100$ MJy sr$^{-1}$ in DF3 and $I_\lambda \sim 300$ MJy sr$^{-1}$ in d203-506. If the infrared radiation field is not isotropic, it is possible its intensity at the H/H$_2$ transition is higher than what we observe in the line of sight. Another estimation of the infrared radiation field at the dissociation front can be made if we consider that it  is mainly emitted by nanograins in the atomic zone. The specific intensity at the dissociation front is then half the intensity of the radiation field in the atomic zone as it is only coming from ahead. In the atomic zone, $I_\lambda \sim 250$ MJy sr$^{-1}$. Thus, both estimation of the infrared radiation field at DF3 are very similar. We then derive  $T_{\rm IR} \sim 150-190$~K in DF3 and d203-506. As a result of different rotational temperatures in the $v=0$ and $v=1$ levels detected in DF3, the vibrational temperature, calculated from Eq. (\ref{eq:Tvib}), varies from $T_{\rm vib} \sim 290$~K to $T_{\rm vib} \sim 650$~K. These temperatures are much higher than that of the near-infrared radiation. 

In addition, as radiative pumping only changes $J$ by $\Delta J = \pm 1$, it is expected that the $v=1$ rotational distribution would mirror the $v=0$ rotational distribution. Figure \ref{fig:diag_rot} shows that the rotational distributions are very different in the $v=0$ and $v=1$, at least for $J>2$. Thus, we conclude that IR pumping is unlikely to account for the excitation of the high-$J$ rotational levels in $v=1$ but also for the first two which cannot be explained by chemical pumping. 

In d203-506, we cannot measure the vibrational temperature between $v=0$ and $v=1$ because there is no data available for the detection of purely rotational lines of CH$^+$. However, the vibrational temperature between $v=1$ and $v=2$ being around $T_{\rm vib} \sim 1300$~K is also incompatible with IR pumping.

Finally, in these environments, collisions are expected to be more important the radiative pumping. Indeed, radiative pumping would dominate over collisions with atomic hydrogen to excite a level $i$ if:

\begin{equation}
   \frac{\lambda^3}{2 h c} \sum_j A_{ij} J_\lambda > \sum_j k_{{\rm coll,} ij} n_{\rm H} 
\end{equation}

However, for all $J$ levels of CH$^+$, $\frac{\lambda^3}{2 h c} \sum_j A_{ij} J_\lambda / \sum_j k_{{\rm coll,} ij} n_{\rm H} \sim~10^{-5}$ in DF3 and $10^{-6}$ in d203-506. This means that radiative pumping does not dominate over collisions with hydrogen. In the next section, we discuss collisions against chemical pumping.

\subsubsection{Collisions}

It is difficult to evaluate accurately the contribution of collisions in the excitation of the $v=1$ state since the inelastic collisional rates between $v=1$ and $v=0$ remain poorly known. Only collisional rates with electrons for the rovibrational excitation have been calculated \citep{Jiang_2019,Forer_2023}. For other collisional partners, only the collisional rates for the rotational levels have been calculated, for atomic hydrogen \citep{Faure_2017} and helium \citep{Hammami_2009}. 

Using the collisional rates with electrons from \cite{Forer_2023}, we find that the collisions with electrons play a negligible role in the excitation of the $v=1$ state compared to chemical pumping in both environments. Indeed, comparing the total excitation rate of the $v=1$ band by collisions and by chemical pumping, collisions with electrons are dominant when 
\begin{equation}
  k_{\text{coll}, v=1} x(e^-) x(\text{CH}^+) > k_{\text{chem}, v=1} x(\text{C}^+) x(\text{H}_2).
  \label{eq:coll_chem}
\end{equation}
where $k_{\text{chem}, v=1}$ is the formation rate of C$^+$ + H$_2$ leading to CH$^+$($v = 1$) and $k_{\text{coll}, v=1}$ is the total collisional excitation rate connecting the $v=0$ and the $v=1$ state. At the H/H$_2$ transition, C$^+$ is the main positive charge carrier so $x(\text{C}^+) \simeq x(e^-)$. Thus Eq. (\ref{eq:coll_chem}) can be rewritten as:
\begin{equation}
   \frac{x(\text{CH}^+)}{x(\text{H}_2)} \gtrsim \frac{k_{\text{chem}, v=1}}{k_{\text{coll}, v=1}} .
  \label{eq:coll_chem_simp}
\end{equation}
\cite{Forer_2023} finds $k_{\text{coll}, v=1}$(1000~K) $\simeq 2 \times 10^{-9}$ cm$^3$ s$^{-1}$ and  $k_{\text{coll}, v=1}$(600~K) $\simeq 2 \times 10^{-10}$ cm$^3$ s$^{-1}$. Quantitatively, we find that collisions become dominant for $x(\text{CH}^+)/x(\text{H}_2) \gtrsim 6 \times 10^{-5}$ in d203-506 and in DF3. According to PDR models $x(\text{CH}^+) \sim  3 \times 10^{-7}$ representative of d203-506 \citep[see Extended Data Fig. 7 from][]{Berne_2023} giving $x(\text{CH}^+)/x(\text{H}_2) \sim 6 \times 10^{-7}$. According to PDR models around $P_{\rm gas} = 5 \times 10^7 - 10^8$~K cm$^{-3}$ and $G_0 = 10^4$, $x(\text{CH}^+) \sim 10^{-7}$ in DF3, thus $x(\text{CH}^+)/x(\text{H}_2) \sim 2 \times 10^{-7}$. In conclusion, collisions with electrons are not dominant in both environments as they would contribute to 1$\%$ at most to the excitation of the $v=1$ state compared with chemical pumping.

However, it is possible that in those environments, collisions with atomic hydrogen are more important than collisions with electrons. Collisions with atomic hydrogen to excite the $v=1$ level are the dominant process when 
\begin{equation}
  k_{\text{coll}, v=1} x(\text{H}) x(\text{CH}^+) > k_{\text{chem}, v=1} x(\text{C}^+) x(\text{H}_2).
  \label{eq:coll_chem_2}
\end{equation}
At the H/H$_2$ transition, $x(\text{H}) \simeq x(\text{H}_2)$. Thus Eq. (\ref{eq:coll_chem_2}) can be rewritten as
\begin{equation}
   \frac{x(\text{CH}^+)}{x(\text{C}^+)} \gtrsim \frac{k_{\text{chem}, v=1}}{k_{\text{coll},v=1}} .
  \label{eq:coll_chem_simp_2}
\end{equation}
This means that chemical pumping tends to be favored if the abundance of the considered species is low compared to the reactants that form it. The collisional rates with H for the rovibrational levels of CH$^+$ are not available. However, using the extrapolation made from \cite{Faure_2017} by \cite{Neufeld_2021}, we can estimate the collisional excitation rate of the $v=1$ state at $T = 1000$~K considering the CH$^+$ rotational population of the $v=0$ level at LTE,  to be $k_{\text{coll}, v=1}(1000 \text{~K}) \simeq 4 \times 10^{-12}$ cm$^3$ s$^{-1}$ and at $T = 600$~K to be $k_{\text{coll}, v=1}(600 \text{~K}) \simeq  3 \times 10^{-13}$ cm$^3$ s$^{-1}$. Thus, collisions become dominant for $x(\text{CH}^+)/x(\text{C}^+) > 5 \times 10^{-2}$ in DF3 and $x(\text{CH}^+)/x(\text{C}^+) > 3 \times 10^{-2}$ in d203-506. Considering $x(\text{C}^+) \simeq 1.4 \times 10^{-4}$, we have $x(\text{CH}^+)/x(\text{C}^+) \sim 10^{-3}$ in DF3 and  $x(\text{CH}^+)/x(\text{C}^+) \sim 3 \times 10^{-3}$ in d203-506. Therefore, we find that collisions with atomic hydrogen would not dominate over chemical pumping in the excitation of the $v=1$ state. However, we observe that collisions with H are more efficient than collisions with electrons and can contribute up to 10$\%$ of the excitation of the $v=1$ band compared with chemical pumping.

Our finding that collision contributes little to the excitation of the $v=1$ state can be seen at odd with the discrepancy between the chemical pumping model and the low-J $v=1$ lines seen in the DF1. 
In fact, up to now, our discussion ignores the rotational distribution of CH$^+$ which should be considered when collisional excitation rates are not negligible compared to the chemical pumping rate. Collisional excitation of the $v=1$ is dominated by collisional transitions starting from the low-J levels of the $v=0$ state since those levels are much more populated than higher-J levels (see excitation diagram Fig. \ref{fig:diag_rot}). As a consequence, the collisions populate preferentially the low-J levels of the $v=1$ due to a combination of the propensity rules ($\Delta J = \pm 1$, $\pm 2$) and the minimization of the energy gap.
In other words, we naturally expect collisions to affect the low-$J$ $v=1$ lines. In contrast, chemical pumping directly produces CH$^+$ within the $v=1$ state in higher $J$-levels. This likely explains why our chemical-pumping model fails only for the lowest $J$ lines within the $v=1$ state by underestimating their emission.
It is likely that in d203-506, CH$^+$ rotational levels within $v=0$ are more evenly populated due to the higher gas temperature and density. This would make the population of the $v=1$ levels by collision more distributed in the different rotational levels and thus less visible than in DF3. Also, our order of magnitude estimates suggest that collisions in d203-506 are less important than in the DF. In any case, the excellent agreement between the chemical-pumping model and the observations of d203-506 remains the most compelling evidence that collisional excitation is negligible in this source for $v=1$ and $v=2$.

We also stress that CH$^+$ is highly reactive and can be destroyed by reaction with H$_2$ and H atoms very rapidly. This high reactivity could hamper the inelastic collision efficiency as an excitation process. In any case, to properly conclude if the $v=1$ $J=0,1$ levels detected in DF3 can be explained by collisions, we need the calculations of the collisional rates of CH$^+$ vibrational bands.

\subsection{Discussion on gas-phase chemistry}

\subsubsection{Role of carbon grain destruction}

Throughout this study, we provide evidence that CH$^+$ is formed and excited by C$^+$ + H$_2$. Indeed, the spatial correlation between H$_2$ and CH$^+$ and the comparison between observed intensities and chemical pumping models shows that both in the Bar and d203-506, the observations can be explained by the gas phase formation route. 

This conclusion is more difficult to make for CH$_3^+$. At the moment, there is no available theoretical or experimental data on the state-resolved reaction CH$_2^+$ + H$_2$. CH$_3^+$ is a 4 atoms system which complicates the modeling of such processes due to the increased number of exit channels. However, several aspects of the observations are striking. CH$_3^+$ is detected at the same position as H$_2$ and CH$^+$. Moreover, in the Bar, its excitation temperature is higher than the gas temperature and higher than the excitation temperature estimated in d203-506, similarly to CH$^+$. Moreover, if CH$^+$ is formed in the gas phase, CH$_3^+$ must be a dominant outcome by the chain reaction of Eq. (\ref{eq:CH3p_form}) because the reactions of CH$^+$ + H$_2$ and CH$_2^+$ + H$_2$ are exothermic for all H$_2(v,J)$ contrary to C$^+$ + H$_2$.

If CH$_3^+$ is not formed in the gas phase, could it be produced by photodestruction of PAHs? Previous studies have shown the importance of this mechanism in small hydrocarbons abundance in the Horsehead Nebula \cite{Pety_2005,Guzman_2015}. Indeed, in this region, state-of-the-art chemical models, which do not take PAH photodestruction process into account, fail to reproduce the abundances of observed hydrocarbons at the PDR edge \citep{Pety_2005,Guzman_2015}. This has been further supported by \cite{Alata_2014,Alata_2015} who showed experimentally that small hydrocarbons can be produced by the photolysis of hydrogenated amorphous carbon when exposed to UV photons. When implementing the experimentally determined reaction parameters in a time-dependent astrophysical model with parameters typical of the Horsehead Nebula, these authors showed that the abundance of small hydrocarbons can temporarily rise by several orders of magnitude. However, these studies do not state the possibility of producing hydrocarbons that are highly hydrogenated such as CH$_3^+$. The methyl radical would have to come from a methyl sidegroup of a PAH and it is more likely that the methyl group would first lose an hydrogen while possibly isomerizing into a tropylium-like structure \citep{Geballe_1989,Joblin_1996,Zhen_2016}. Studies on PAH photodestruction mainly reveal the formation of hydrocarbons with small carbon chains containing at least two carbons and no ionization is expected from neutral fragmentation. To this date, there is still no detection of the nonpolar species CH$_4$ and C$_2$H$_2$ in the Orion Bar which are expected to be the main products of the photodestruction of PAHs or amorphous carbon grains. In addition, \cite{Cuadrado_2015} observed a variety of small hydrocarbons in this region with the IRAM 30m telescope and has been able to explain their abundances only with gas phase chemistry \citep[see also][]{Goicoechea_2025}. Finally, it is also hard to conceive that CH$_3^+$ could be produced with such an excitation temperature in the Bar through the photodestruction of small grains. 

All these results are in favor of the hypothesis that CH$_3^+$ is formed and chemically pumped in the gas phase (see Eq. (\ref{eq:CH3p_form})) under the conditions explored here in Orion. State-to-state rate coefficients of the chemical pumping of CH$_3^+$ through the reaction CH$_2^+$ + H$_2$ are essential to properly conclude this matter.

\subsubsection{Searching for CH$_2^+$}

In the gas-phase route, we thus detect two crucial intermediates: CH$^+$ and CH$_3^+$. Assuming this route predominantly forms CH$_3^+$ (see Eq. (\ref{eq:CH3p_form})), the question arises: what happens to CH$_2^+$, the intermediate step between CH$^+$ and CH$_3^+$? CH$_2^+$ possesses three vibrational modes, the symmetric stretching mode ($\nu_1$), the bending mode ($\nu_2$), and the asymmetric stretching mode ($\nu_3$) and their corresponding fundamentals in its electronic ground state are located around $\sim$3.47~$\upmu$m (2883.0 cm$^{-1}$) \citep{Kraemer1994}, $\sim$10.05~$\upmu$m (995.5 cm$^{-1}$) \citep{Kraemer1994}, and $\sim$3.19~$\upmu$m (3131.4 cm$^{-1}$) \citep{Rosslein1992}, respectively. Following the logic behind the detection of CH$^+$ and CH$_3^+$, one would expect the detection of CH$_2^+$ based on the emission from $\nu_2$ around 10~$\upmu$m supposed to be the brightest. The existing experimental data in the literature do not allow us to currently model a high-resolution gas phase emission spectrum for this band due to the lack of spectroscopic information about the rotational constants of the $\nu_2=1$ vibrational state. Only the rotational constants of the ground state are well characterized \citep{Rosslein1992, Willitsch2003, Gottfried2004}. The fundamental band of this mode has been reported theoretically \citep{Osmann_1997} and experimentally \citep{Bunker2001}. These results constrain the search for this emission feature from CH$_2^+$ between 9 and 11~$\upmu$m, a region where there are no remaining unassigned observed transitions in the JWST spectra of DF3 and d203-506 (see Fig. \ref{fig:ch2+}). The search of CH$_2^+$ signatures in the $\nu_1$ and $\nu_3$ band in the NIR was also unsuccessful. In addition, no unidentified lines in the ISO and \textit{Herschel}/PACS spectrum can be assigned to CH$_2^+$ rotational emission. This raises the intriguing issue as to why CH$^+$ and CH$_3^+$ can be detected by their emission fingerprints while CH$_2^+$ cannot, despite astrochemical models forecasting comparable abundances for CH$_2^+$ and CH$_3^+$ \citep[see Extended Data Fig. 7 of][]{Berne_2023}. \cite{Mazo-Sevillano_2024} predicts that CH$_2^+$ is slightly less abundant than CH$_3^+$ which could explain its non detection, if it is not chemically pumped, in the Bar but not in d203-506.

\begin{figure}
\begin{center}
\includegraphics[width=\columnwidth,angle=0]{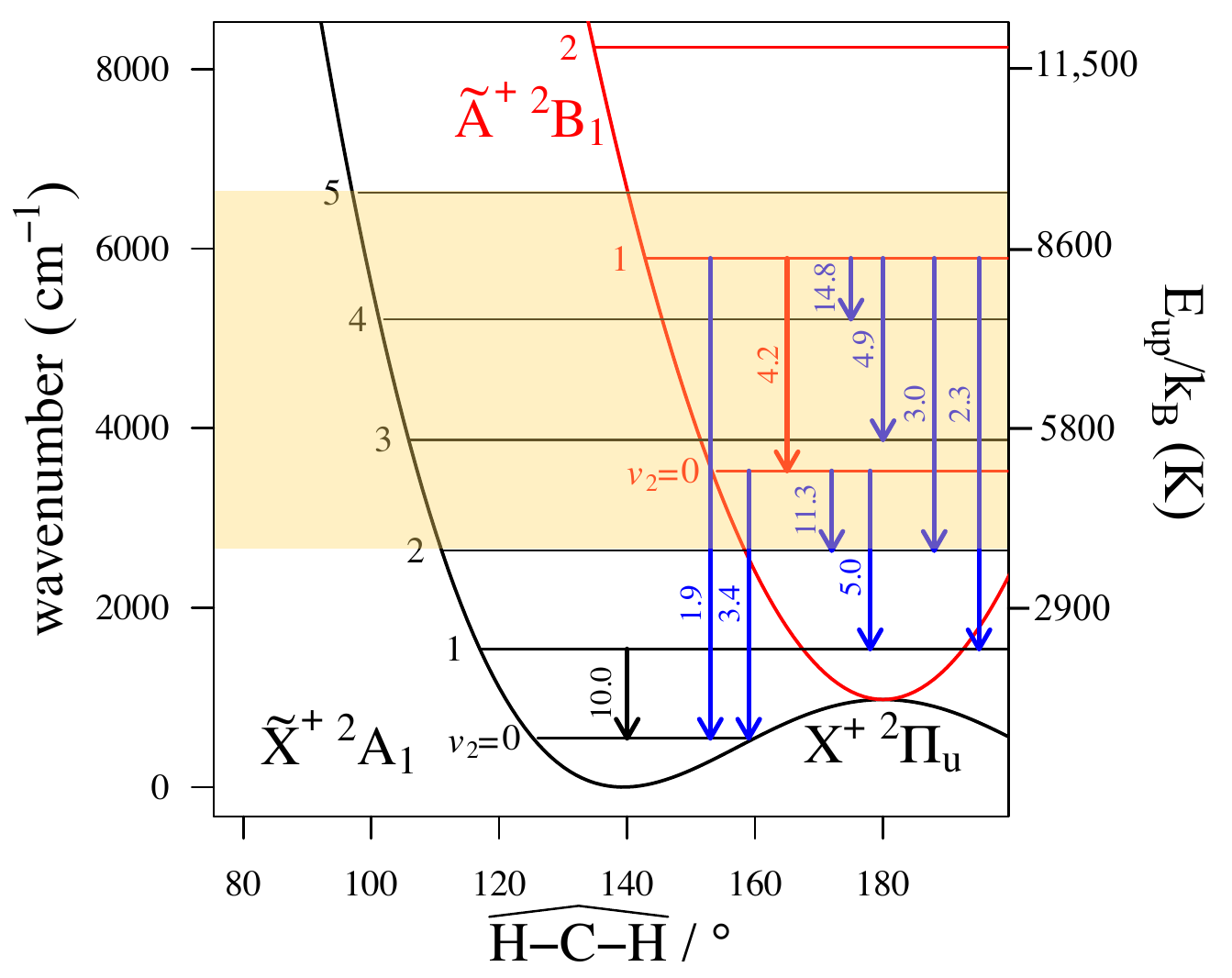}
\caption{Energy level diagram of the bending levels in the two lowest electronic states of CH$_2^+$, shown as a function of the HCH bond angle in degrees, with no quanta of excitation in either stretching mode. These two electronic states are strongly coupled by Renner-Teller effect. The allowed vibrational (black for the fundamental $\tilde{\mathrm{X}}$ state, red for the $\tilde{\mathrm{A}}$ state) and vibronic (blue) transitions are shown. 
Curves of potential and vibrational levels positions are adapted from \cite{Coudert2018} and \cite{Jensen1995}. For each transition, the position they will fall in is labeled in~$\upmu$m. The box overlayed in yellow represents the upper energy levels of detected CH$^+$ transitions.}
\label{Figure_niveaux_ch2_plus}
\end{center}
\end{figure}

\begin{figure*}
    \centering
    \includegraphics[width=\linewidth]{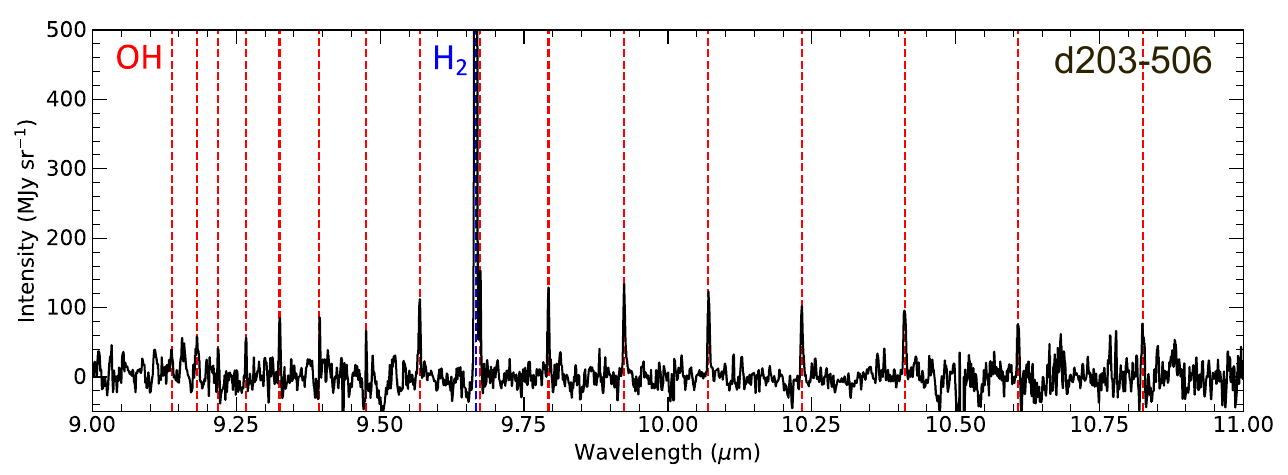}
    \caption{Continuum and OFF position subtracted spectrum of the disk d203-506 between 9 and 11~$\upmu$m with the identified lines. The OH lines highlighted in red were previously presented and analyzed in \cite{Zannese_2024}.}
    \label{fig:ch2+}
\end{figure*}

A better understanding of CH$_2^+$ spectroscopic properties is required to assess how compatible this non-detection is with current predictions by astrochemical models. The initial step to answer this question involves describing the strong vibronic couplings known to be present in CH$_2^+$ bending modes and giving rise to a rich manifold of potential transitions (see Fig. \ref{Figure_niveaux_ch2_plus}) and assessing the population distribution of $\rm CH_2^+$ in the energy level structure during the process of its formation from the addition of H$_2$ to CH$^+$. In the absence of specific information on the state distribution of $\rm CH_2^+$ in the Orion Bar, the formation of $\rm CH_2^+$ in its first electronic excited state ($\rm \tilde{A}^+$) has also to be considered. $\rm CH_2^+$ could present vibrational emission from its least energetic vibrational mode ($\rm\nu_2=1 \rightarrow \nu_2=0$ in $\rm \tilde{A}^+$), which should be located around $\sim$ 4~$\upmu$m (2500 cm$^{-1}$) . Alternatively, it could relax to its ground electronic state ($\rm \tilde{X}^+$) by vibronic emission. If one estimates that only the first vibrational levels of the $\rm \tilde{A}^+$ state are populated based on CH$^+$ and CH$_3^+$ temperatures in the Orion Bar (estimated to lie between about 700 and 1500~K), the strongest emission transitions will be towards the excited vibrational levels of $\rm \tilde{X}^+$. Emission features of $\rm CH_2^+$ could therefore originate from the $\rm \nu_2=0$ and $\rm \nu_2=1$ of the $\rm \tilde{A}^+$ electronic state towards the $\rm \nu_2$ bending levels of the $\rm \tilde{X}^+$ electronic state \citep{Wang_2013,Bunker_2007}. Expected ranges for such transitions are reported in Fig~\ref{Figure_niveaux_ch2_plus}. Moreover, \cite{Bunker_2007} finds that the strongest of these bands are in the near-infrared range around 0.9 $\mu$m. Typical vibronic emission has a larger probability than vibrational emission, which could explain the elusiveness of $\rm CH_2^+$ as its spectroscopic signatures could be spread and contribute in another spectral region of the JWST spectra. The emission spectrum of $\rm CH_2^+$ in the hot environment of the Orion Bar and d203-506 could actually be located in the NIR region, especially if $\rm CH_2^+$ is produced in vibrational states higher than $\rm \nu_2=1$ in the $\rm \tilde{A}^+$ electronic state. Some lines are still unidentified in the JWST NIR spectrum \citep{Peeters_2024} but the identification to CH$_2^+$ has not been made. The precise calculation of CH$_2^+$ spectroscopy is necessary to compare the strength of the different possible transitions ($\nu_1, \nu_2, \nu_3$) and conclude on its non-detection. Moreover, further investigation of $\rm CH_2^+$ formation pathways and degree of excitation are required to estimate the population of the $\tilde{A}$ state in order to evaluate the importance of rovibronic transitions. However, due to the complexity of the calculations, this is referred to a future study.\\

\subsubsection{What happens after CH$_3^+$?}

CH$_3^+$ is hypothesized to be at the origin of most of the complex organic chemistry in the UV-irradiated gas. By reacting with small hydrocarbons, CH$_3^+$ can produce long-chain hydrocarbons. A variety of hydrocarbons are indeed detected with the IRAM 30m telescope, \textit{Herschel} and ALMA in the Orion Bar \citep{Cuadrado_2015, Nagy_2015,Goicoechea_2025}. Interestingly, CH$_3^+$ is also ont of the origin of very abundant molecules such as HCO$^+$ which is the direct product of the reaction:

\begin{equation}
        \label{eq:hco+}
        \ce{CH_3+ + O <=> HCO^+ + H$_2$} 
\end{equation}

As the atomic oxygen is very abundant, this reaction is as important as the formation of HCO$^+$ from CO$^+$ + H$_2$ = HCO$^+$ + H as predicted by PDR models \citep[see Extended Data Fig. 7 and 8 of][]{Berne_2023}. Thus CH$_3^+$ is an important source of CO as well. This is compatible with previous studies that have shown the spatial coincidence between the rotational lines of HCO$^+$ and the rovibrational emission of H$_2$ in the Orion Bar \citep{Goicoechea_2016, Habart_2023} and recent observations with ALMA of HCO$^+$ $J=4-3$ presented in \cite{Berne_2024} which follows quite well the H$_2$ and CH$_3^+$ emission in. Interestingly, a tentative signature of the Q-branch of HCO$^+$ is found around 12~$\upmu$m in d203-506 with MIRI-MRS (see Fig. \ref{fig:hco+}), which would be the IR counterpart of the rotational emission detected by ALMA and the third species detected by JWST in this reaction chain. The analysis of this signature is beyond the scope of this paper and referred to futur investigations. In any case, our analysis reveals the first steps of the chemical route that starts from C$^+$ and opens new avenues to consistently analyze the emission of other species detected through their rotational or rovibrational emissions.

\section{Conclusion}

In this work, we study the rovibrational emission of CH$^+$ and CH$_3^+$ in the Orion Bar and establish constraints on their formation route and excitation. We also explored the potential of NIR CH$^+$ emission as a diagnostic for the study of interstellar PDRs and disks. 

The main conclusions of this study can be summarized as follows:

\begin{enumerate}
    \item CH$^+$, CH$_3^+$ rovibrational emission and excited H$_2$ emission originate from the same regions. They trace a thin layer of the H$^0$/H$_2$ transition where the emission from FUV-pumped H$_2$ levels peaks too.\\

    \item CH$^+$ and CH$_3^+$ both have higher excitation temperature in the Bar than the gas temperature ($T_{\rm ex}$(Bar) $ \sim 1500$~K 
    $> T_{\rm gas}$(Bar) $ \sim 570$~K). In addition, they have higher excitation temperature in the Bar than in d203-506, even if the gas temperature in the disk is higher ($T_{\rm ex}$(d203-506) $ \sim 850$~K $\sim$ $T_{\rm gas}$(d203-506) $ \sim 850$~K). \\

    \item The study of the observed column densities of CH$^+$ and CH$_3^+$ in their rovibrational states shows that the observed emission traces only a small fraction of the total abundance of these molecules, less than 0.1 \% in the Bar. Indeed, for both molecules, the column density derived from the rovibrational lines is around $N_{\rm vib}$(CH$^+$/CH$_3^+$) $\sim 10^{10}$ cm$^{-2}$, when the total expected column densities are around $N$(CH$^+$/CH$_3^+$) $\sim 10^{14}-10^{15}$ cm$^{-2}$.\\
    
    \item The excitation of CH$^+$ and probably CH$_3^+$ can be explained by chemical formation pumping with excited H$_2$. This is in favor of a formation route in the gas phase following \ce{C^+ ->[H_2] CH^+ ->[H_2] CH_2^+ ->[H_2] CH_3^+}.\\

    \item The in-depth study of the chemical formation pumping mechanism has shown that in regions where the temperature and density are high (such as d203-506), the excitation of CH$^+$ is mostly driven by H$_2$ rotational levels which are populated by collisions. In contrast, in regions where the temperature is not as high and the gas is less dense but highly irradiated (such as the Bar), the excitation is mostly driven by FUV-pumped levels of H$_2$. \\

       \item Chemical formation pumping models can be used to determine the formation rate of CH$^+$ in the Orion Bar and in d203-506. \\

        \item CH$^+$ and CH$_3^+$ emission strength depends strongly on the local density. As the excitation of CH$^+$ follows a non-thermal process, a simple diagnostic gives access to the gas density at the H/H$_2$ transition. This diagnostic provides an estimate of $n_{\rm H} \simeq 10^7$ cm$^{-3}$ in d203-506 and $n_{\rm H} \simeq 10^6$ cm$^{-3}$ in the Bar. The latter might be overestimated due to the use of large apertures to derive line intensities.   \\

    \item Attempts to detect CH$_2^+$, the intermediate between CH$^+$ and CH$_3^+$, were unsuccessful. However, the complexity of CH$_2^+$ spectroscopy could explain its non-detection even if it has similar abundances to CH$^+$ and CH$_3^+$.

\end{enumerate}

In conclusion, our study has unveiled the gas phase route of carbonaceous species. However, it also revealed the need for molecular data to be able to properly interpret the JWST observations. Most of this study relies on extrapolated rates which does not allow us to compare the chemical pumping rates coming from vibrational and rotational levels of H$_2$ with similar energies. Moreover, while chemical formation pumping is thoroughly studied for CH$^+$, the state-to-state rate coefficients are not available for CH$_3^+$. However, they are essential to understand the formation and excitation process of this molecule.

\begin{acknowledgement}
 We would like to thank I. Schneider for our fruitful discussion on collisions between CH$^+$ and $e^-$ which helped us discuss this matter. This work is based [in part] on observations made with the NASA/ESA/CSA James Webb Space Telescope. The data were obtained from the Mikulski Archive for Space Telescopes at the Space Telescope Science Institute, which is operated by the Association of Universities for Research in Astronomy, Inc., under NASA contract NAS 5-03127 for JWST. These observations are associated with program ERS1288. This work was partially supported by the Programme National “Physique et Chimie du Milieu Interstellaire” (PCMI) of CNRS/INSU with INC/INP.  This work was supported by CNES with funds focused on JWST. J.R.G. thanks the Spanish MCINN for funding support under grant PID2023-146667NB-I00. E.P. and J.C. acknowledge support from the University of Western Ontario, the Institute for Earth and Space Exploration, the Canadian Space Agency (CSA, 22JWGO1- 16), and the Natural Sciences and Engineering Research Council of Canada. This article is based upon work from COST Action CA21126 - Carbon molecular nanostructures in space (NanoSpace), supported by COST (European Cooperation in Science and Technology).
\end{acknowledgement}

\bibliographystyle{aa}

\begin{appendix}

\section{Tentative detection of HCO$^+$ in MIRI-MRS}

\begin{figure}[!h]
    \centering
    \includegraphics[width=\linewidth]{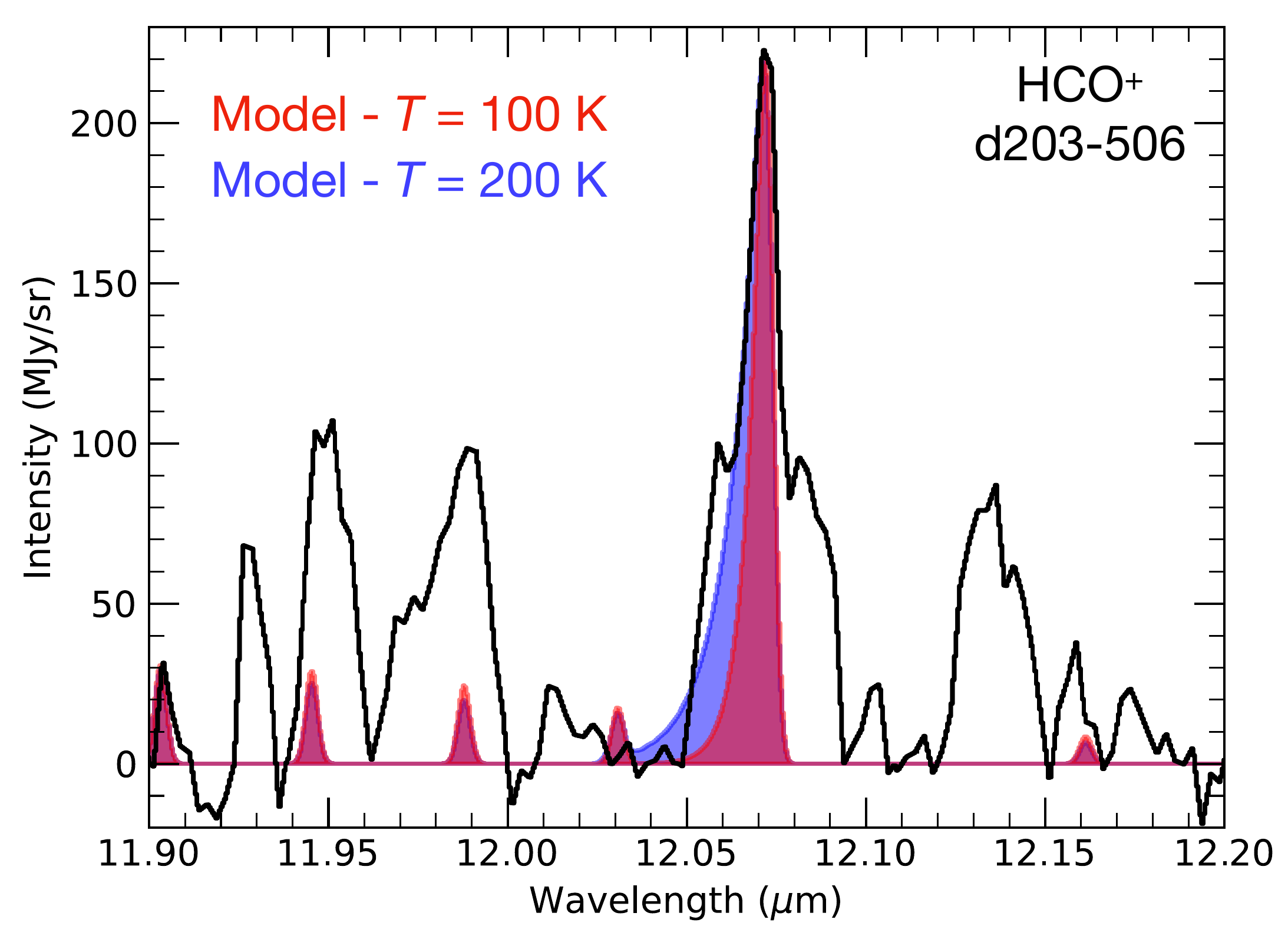}
    \caption{Tentative detection of HCO$^+$ in the disk d203-506 with MIRI-MRS. Two models at $T=100$~K (in red) and $T=200$~K (in blue) using spectroscopic data from \cite{Davies_1984,Foster_1984} are superposed to the observations.}
    \label{fig:hco+}
\end{figure}

A tentative detection in MIRI-MRS of HCO$^+$ is presented in Fig. \ref{fig:hco+}. The comparison to simulated spectra at $T=100$~K and $T=200$~K shows rather cold emission of HCO$^+$. The exact excitation of HCO$^+$ needs further investigation. The excitation temperature is lower than the gas temperature questions pointing toward subthermal excitation of the rotational levels, which might be explained by IR pumping. However the proper analysis of this signature is referred to a future study. 

\section{Importance of non-LTE distribution of H$_2$ in chemical pumping}

\begin{figure}
    \centering
    \includegraphics[width=\linewidth]{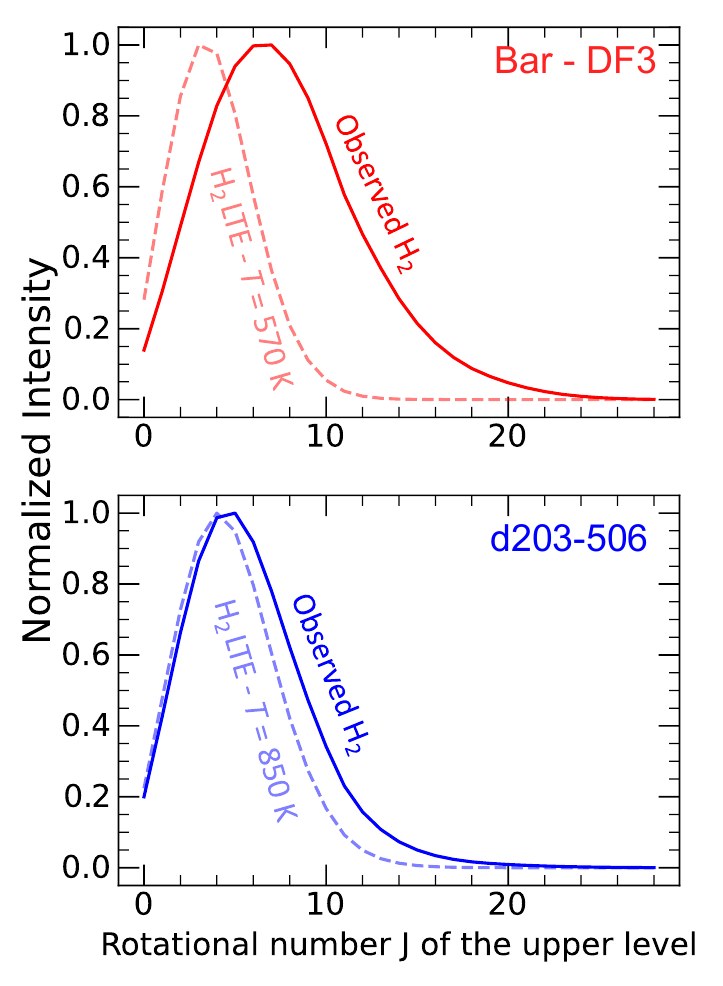}
    \caption{Comparison of modeled chemical pumping intensities of CH$^+$ using LTE distribution of H$_2$ at the gas temperature and the observed distribution in the Orion Bar. (Top) Modeled intensities for DF3. (Bottom) Modeled intensities for d203-506. Taking into account FUV-pumped levels of H$_2$ is highly important in DF3 compared to d203-506.}
    \label{fig:distrib_comp_ETL}
\end{figure}

Figure \ref{fig:distrib_comp_ETL} shows that taking into account the observed non-LTE distribution of H$_2$ in the CH$^+$ chemical models is highly important for DF3. In d203-506, the modeled intensities for an H$_2$ LTE distribution and the observed distribution is very similar. This highlights the importance of thermalized levels of H$_2$ in the excitation of CH$^+$ in the disk. Thus, we find that in environments with high temperature, pure rotational levels of H$_2$ are sufficient to excite CH$^+$ in the $v=1$ level. When the temperature is lower, the FUV-pumped levels of H$_2$ have higher importance in the excitation.

\section{CH$^+$ line intensities in the Bar and in d203-506}
Table \ref{tab:intensity_chp} presents the CH$^+$ line intensities detected in DF3 not and corrected for extinction and in d203-506.

\begin{table*}[!h]
    \centering
    \begin{tabular}{c|c|c|c|c|c}
        Wavelength ($\upmu$m) & Line & E$_{up}$/k$_B$ (K) & Intensity - DF3 & Intensity - DF3 &  Intensity - Disk   \\
        &  &  & ($\times 10^{-6}$ erg cm$^{-2}$ s$^{-1}$ sr$^{-1}$) & (corrected for extinction)  & ($\times 10^{-5}$ erg cm$^{-2}$ s$^{-1}$ sr$^{-1}$)  \\
         &  &  &  & ($\times 10^{-6}$ erg cm$^{-2}$ s$^{-1}$ sr$^{-1}$)  &  \\
        \hline
        \multicolumn{6}{c}{$v=1 \rightarrow 0 $} \\
        \hline
        3.549 & R(3) & 4328  & 1.51 $\pm$ 0.49 & 1.94 $\pm$ 0.67    & 1.14 $\pm$ 0.26 \\   
        3.581 & R(2) & 4174  & 1.97 $\pm$ 0.88  & 2.53 $\pm$ 1.13  & 0.71 $\pm$  0.26 \\
        3.615 & R(1) & 4058 & 0.59 $\pm$ 0.19 & 0.76 $\pm$ 0.24  & 1.27 $\pm$ 0.28 \\
        3.688 & P(1) & 3942 & 4.34 $\pm$ 0.89 & 5.52 $\pm$ 1.13 & 1.09 $\pm$ 0.28 \\ 
        3.727 & P(2) & 3980 & 3.87 $\pm$ 0.28 &  4.90 $\pm$ 0.35 & 2.30 $\pm$ 0.25 \\
        3.769 & P(3) & 4058 & 3.03 $\pm$ 0.65 & 3.83  $\pm$ 0.82 & 3.64 $\pm$ 0.16 \\
        3.813 & P(4) & 4174 & 4.13 $\pm$ 1.11 & 5.20 $\pm$ 1.40 & 4.17 $\pm$ 0.09 \\
        3.859 & P(5) & 4328 & 3.96 $\pm$ 1.17 & 4.97 $\pm$ 1.47 & 5.16 $\pm$ 0.17 \\
        3.959 & P(7) & 4751 & 5.02 $\pm$ 0.43 & 6.26 $\pm$ 0.53  & 4.56 $\pm$ 0.19 \\
        4.013 & P(8) & 5019 & 6.28 $\pm$ 0.66 & 7.79 $\pm$ 0.81 & 4.36 $\pm$ 0.27 \\
        4.128 & P(10) &  5324 & 4.53 $\pm$ 1.85& 5.58 $\pm$ 1.78 & / \\
        4.190 & P(11) &  6046 & 4.33 $\pm$ 1.05 & 5.32 $\pm$ 1.29 & 1.82 $\pm$ 0.22 \\
        4.255 & P(12) &  6461& 2.81 $\pm$ 0.84 & 3.44 $\pm$ 1.03 & 1.46 $\pm$ 0.14 \\
        4.323 & P(13) &  6912& 3.63 $\pm$ 0.41 & 4.43 $\pm$ 0.50 & 1.13 $\pm$ 0.07 \\
        4.394 & P(14) & 7398 & / & / & 0.84 $\pm$ 0.21 \\      \hline
        \multicolumn{6}{c}{$v=2 \rightarrow 1$} \\  
        \hline
        3.570 & R(7) & 9048 & / &/& 1.12 $\pm$ 0.84 \\
        3.676 & R(3) & 8088& / & / & 0.12 $\pm$ 0.01 \\	
        3.707& R(2) &  7939 & / & / & 0.38 $\pm$ 0.06 \\
        4.195 & P(8) & 8754 & / &/ & 0.48 $\pm$ 0.16	\\
        4.318 & P(10) &9378& /& /& 0.50 $\pm$ 0.08 \\
        4.383 & P(11) &  9743 & /&/ & 0.27 $\pm$ 0.09 \\
    \end{tabular}
    \caption{Intensities of the CH$^+$ lines detected in DF3 and in d203-506. The uncertainties are only the fitting error. The impact of calibration effects is not taken into account and the associated uncertainties can be as high as 20\%.}
    \label{tab:intensity_chp}
\end{table*}

\section{Observed spectra of CH$^+$ and CH$_3^+$ in the Bar and in d203-506}
\subsection{CH$^+$}
\begin{figure*}[!h]
	    \centering
     	    \includegraphics[width=\linewidth]{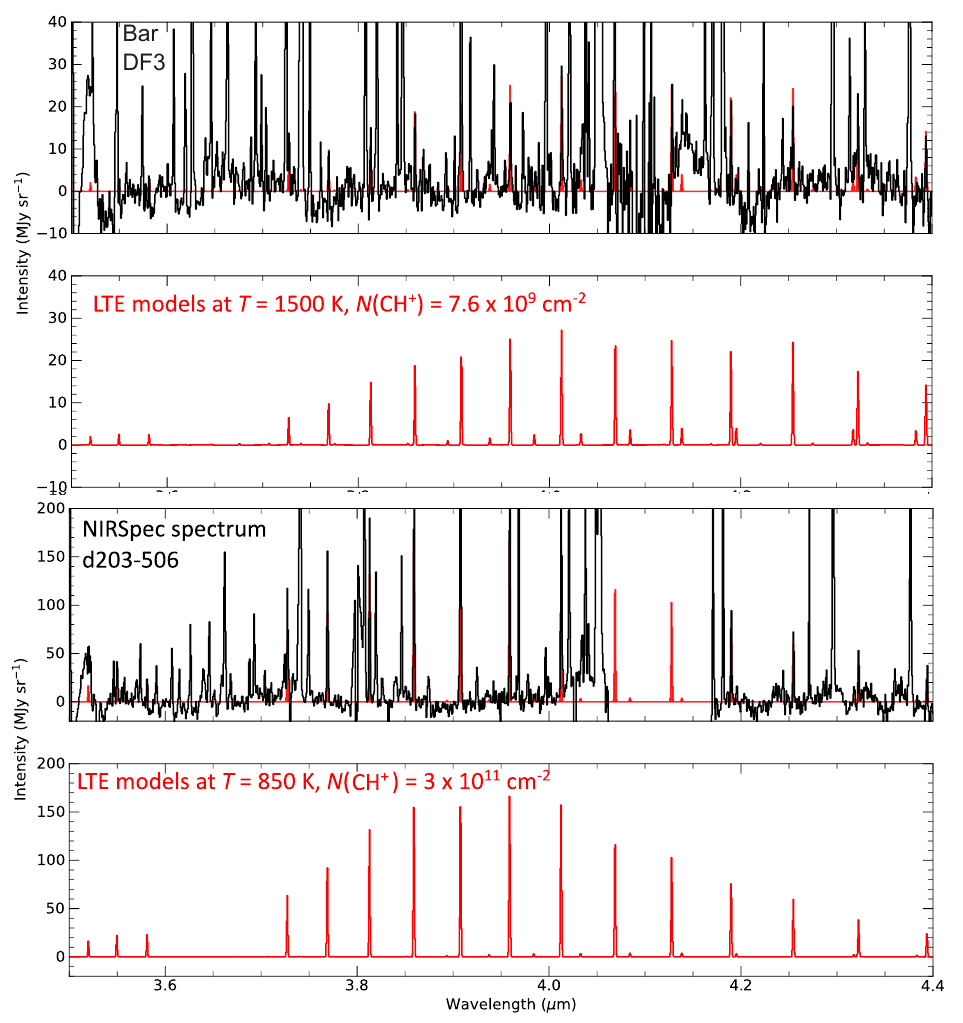}
	    \caption{(a) (Top) CH$^+$ continuum subtracted spectra in DF3 (see Fig. \ref{fig:OB_RGB}) (Bottom) LTE model of CH$^+$ at $T=1500$~K and $N_{\rm vib}$(CH$^+$) $=7.6 \times 10^9$ cm$^{-2}$. The column density of the fit is lower due to extinction in the line of sight. (b) CH$^+$ continuum subtracted spectra in d203-506 (see Fig. \ref{fig:OB_RGB}) (Bottom) LTE model of CH$^+$ at $T$=850K and $N_{\rm vib}$(CH$^+$) $= 3 \times10^{11}$ cm$^{-2}$.}
	    \label{fig:chp_spectra}
	\end{figure*}

Fig. \ref{fig:chp_spectra} displays the observed spectra of CH$^+$ in DF3 and d203-506 with fitted LTE models.

\subsection{CH$_3^+$}
\begin{figure*}[!h]
    \centering
    \includegraphics[width=\linewidth]{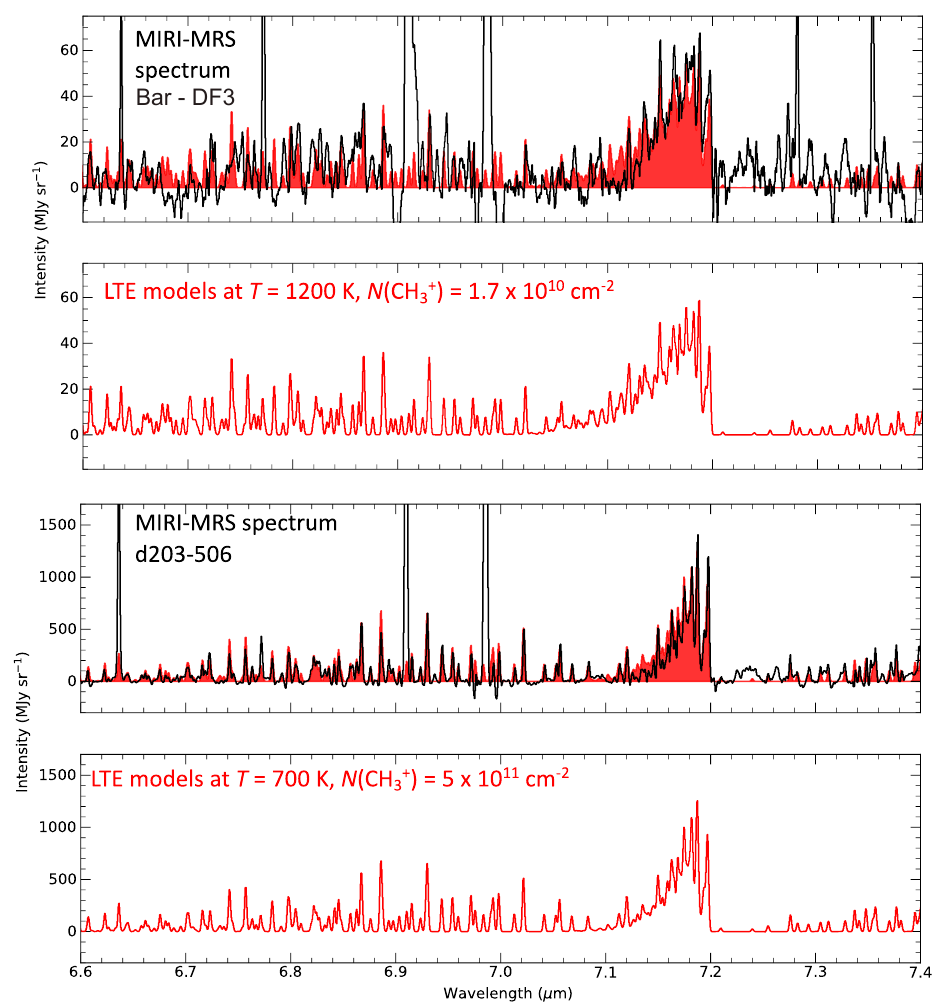}
    \caption{(a) (Top) CH$_3^+$ continuum subtracted spectra in DF3 (see Fig. \ref{fig:OB_RGB}). (Bottom) LTE model of CH$_3^+$ at $T=1200$~K and $N_{\rm vib}$(CH$_3^+$) $=1.7 \times 10^{10}$ cm$^{-2}$. (b) (Top) CH$_3^+$ continuum subtracted spectra in d203-506 (see Fig. \ref{fig:OB_RGB}). (Bottom) LTE model of CH$_3^+$ at $T=700$~K and $N_{\rm vib}$(CH$_3^+$) $=5 \times 10^{11}$ cm$^{-2}$ \citep{Berne_2023,Changala_2023}.}
    \label{fig:ch3p_spectra}
\end{figure*}

Fig. \ref{fig:ch3p_spectra} displays the observed spectra of CH$_3^+$ in DF3 and d203-506 with fitted LTE models.

\section{Dependence of chemical pumping rates of CH$^+$ ($v=1$) on H$_2$ levels ($v',J'$)}

\begin{figure*}
        \centering
        \includegraphics[width=1\linewidth]{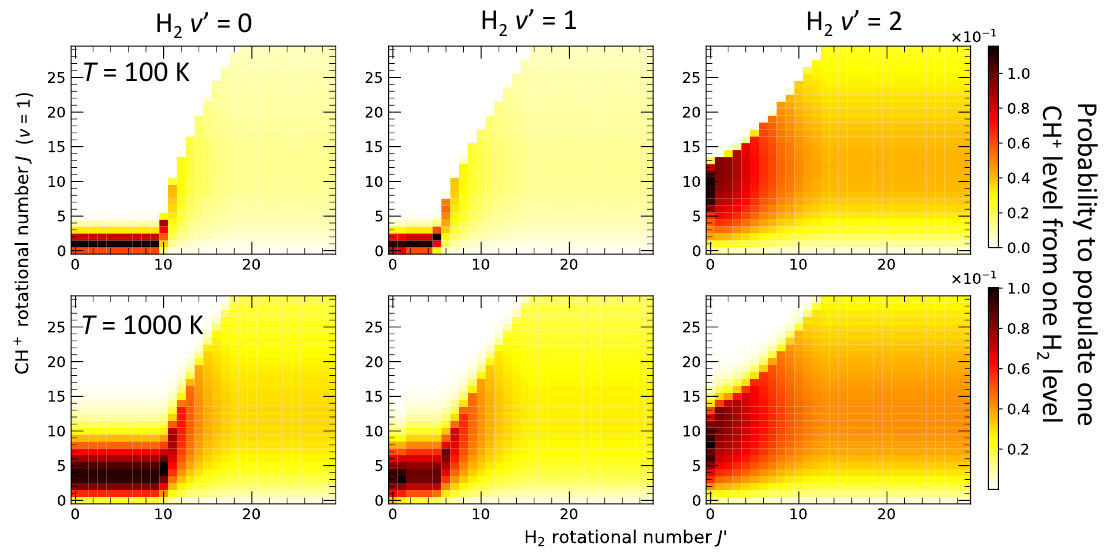}
        \caption{Probability to populate one rotational level of CH$^+$ ($v=1$) from one level of H$_2$. Only the chemical pumping rates from the $v'=1$ $J'=0,1$ and $v'=2$ $J'=0$ levels of H$_2$ are available \citep{Zanchet_2013,Faure_2017}. The other rates come from the extrapolation made in \cite{Neufeld_2021}. Two regimes can be observed. The first and flat one ($J' < 10$ for $v'=0$) is when $E({\rm H}_2) < E(\text{CH}^+)+\Delta E$. The second one, cone-like ($J' > 10$ for $v'=0$), is when $E({\rm H}_2) > E(\text{CH}^+)+\Delta E$ and the chemical pumping rate depends neither on the temperature nor on the levels energy of CH$^+$ and H$_2$ anymore.}
        \label{fig:proba_chp_h2}

\end{figure*}

Figure \ref{fig:proba_chp_h2} shows the dependence of chemical pumping rates of CH$^+$ ($v=1$) on different H$_2$ levels and on gas temperature. We find that H$_2$ levels with high excitation are able to populate higher-$J$ levels of CH$^+$. When $E({\rm H}_2) > E(\text{CH}^+)+\Delta E$, the rate coefficients depends neither on the temperature nor on the levels energy of CH$^+$ and H$_2$ anymore.

\section{Post-processing of the JWST spectra}
\subsection{CH$^+$}
\begin{figure*}[!h]
    \centering
    \includegraphics[width=\linewidth]{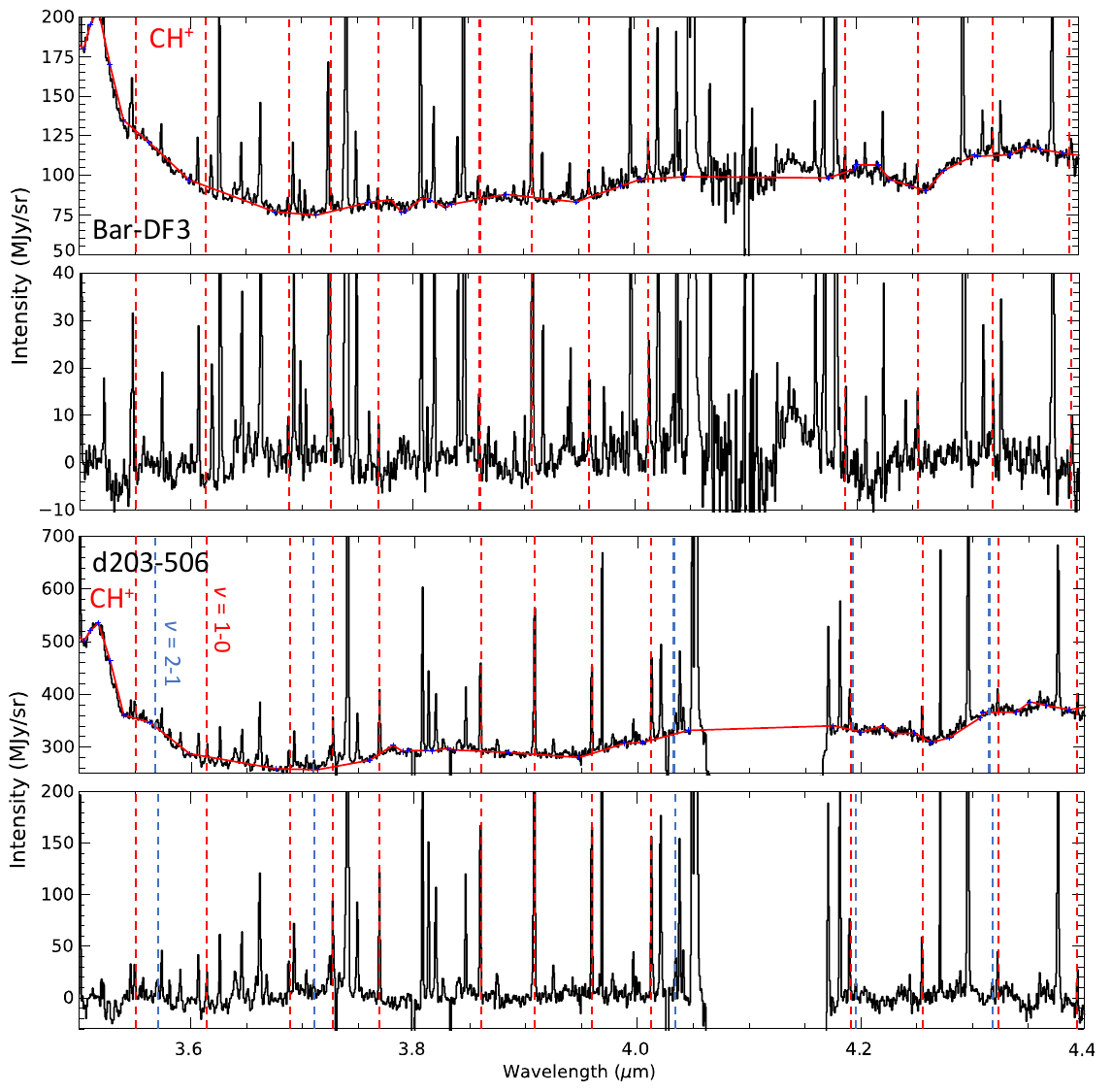}
    \caption{Processing of the NIRSpec spectrum. (a) DF3 : (Top panel) Spectrum observed with NIRSpec. The red line is the estimated continuum. (Bottom panel) Continuum
subtracted spectrum. (b) Disk (Top panel) Spectrum observed with NIRSpec. The red line is the estimated continuum. (Bottom panel) Continuum
subtracted spectrum. Red vertical lines are CH$^+$ $v=1-0$ transitions and blue lines are CH$^+$ $v=2-1$ transitions. Other line identifications can be found in \cite{Peeters_2024}}.
    \label{fig:process_chp}
\end{figure*}

Fig. \ref{fig:process_chp} displays the post-processing of the NIRSpec observed spectra to better visualize CH$^+$ lines in DF3 and d203-506.

\subsection{CH$_3^+$}
\begin{figure*}[!h]
    \centering
    \includegraphics[width=\linewidth]{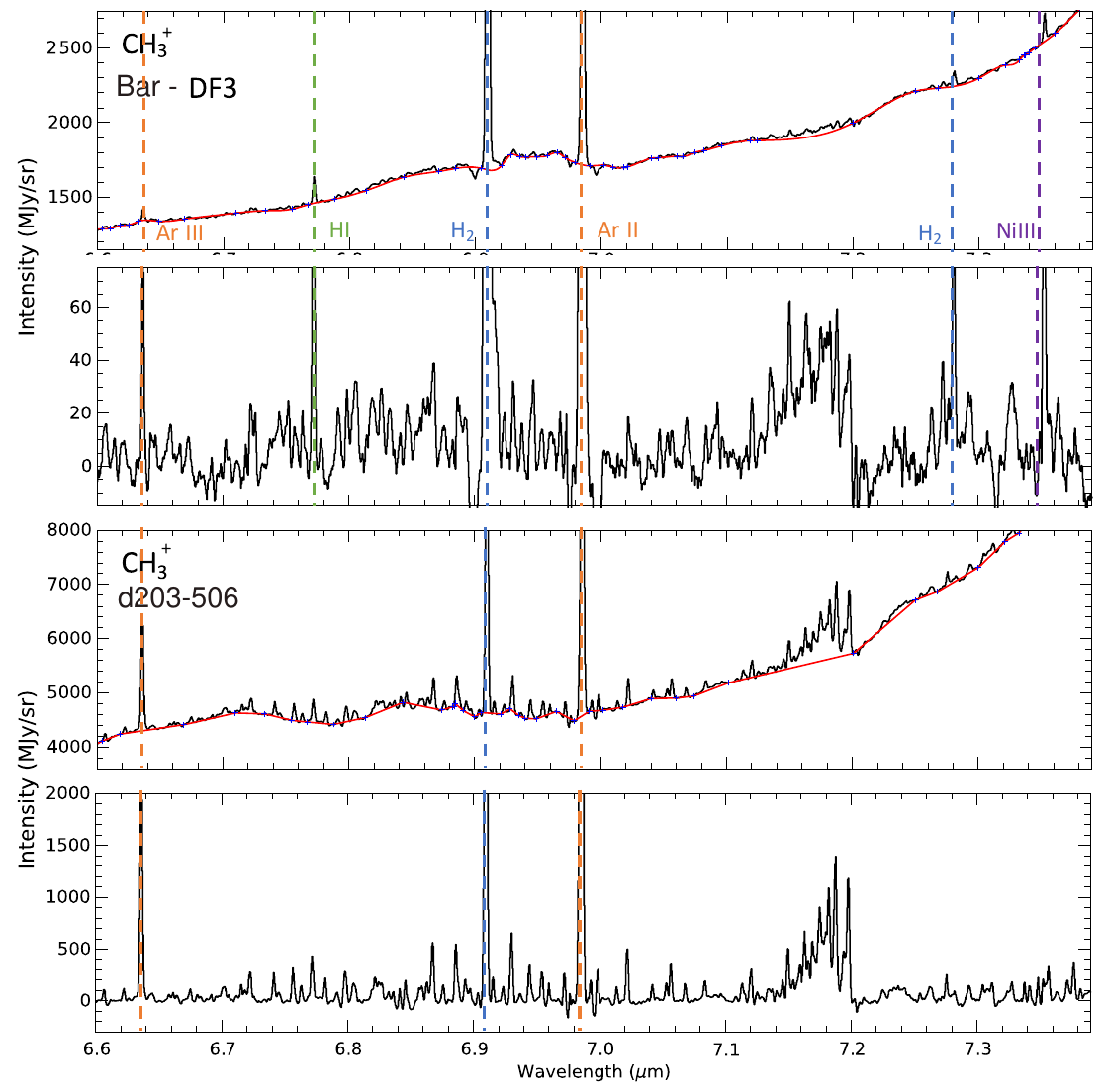}     
    \caption{Processing of the MIRI-MRS spectrum. (a) DF3 : (Top panel) Spectrum observed with MIRI-MRS. The red line is the estimated continuum. (Bottom panel) Continuum subtracted spectrum. (b) Disk : (Top panel) Spectrum observed with MIRI-MRS. The red line is the estimated continuum. (Bottom panel) Continuum
subtracted spectrum. All the observed lines are from CH$_3^+$ except for the highlighted ones. The identifications are made using \cite{van_de_putte_2024}.}
    \label{fig:process_ch3p}
\end{figure*}

Fig. \ref{fig:process_ch3p} displays the post-processing of the MIRI-MRS observed spectra to better visualize CH$_3^+$ feature in DF3 and d203-506.
\subsection{HCO$^+$}
\begin{figure*}[!h]
    \centering  
    \includegraphics[width=\linewidth]{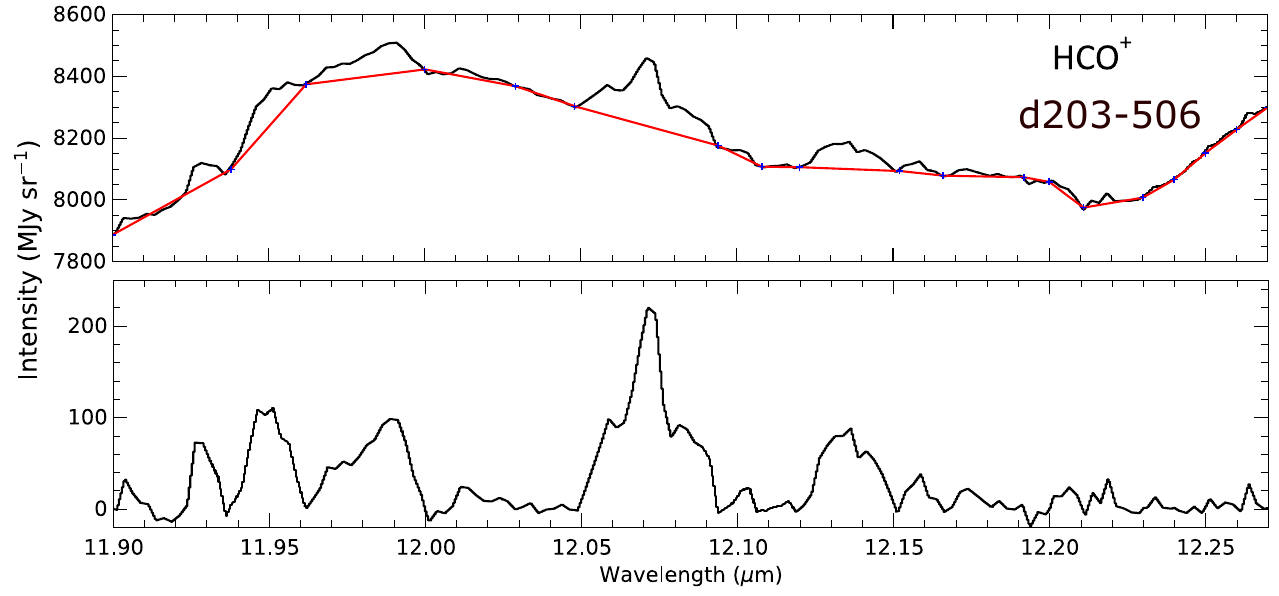}    
    \caption{Processing of the MIRI spectrum. (Top panel) Spectrum observed with MIRI. The red line is the estimated continuum. (Bottom panel) Continuum subtracted spectrum. }
    \label{fig:process_hcop}
\end{figure*}

Fig. \ref{fig:process_hcop} displays the post-processing of the MIRI-MRS observed spectra to better visualize HCO$^+$ feature in d203-506.

\end{appendix}
\end{document}